\documentclass[11pt]{article}
\usepackage{graphicx,amssymb,amsbsy}

\begin{document}

\title{Multi-Planet Destabilisation and Escape in Post-Main Sequence Systems}
\author{George Voyatzis$^{1}$\thanks{voyatzis@auth.gr},
John D. Hadjidemetriou$^{1}$\thanks{E-mail:hadjidem@auth.gr},
Dimitri Veras$^{2}$\thanks{E-mail:veras@ast.cam.ac.uk}, and
Harry Varvoglis$^{1}$\thanks{E-mail:varvogli@physics.auth.gr}  
\\
$^{1}$Department of Physics, University of Thessaloniki, GR-541 24 Thessaloniki, Greece 
\\
$^{2}$Institute of Astronomy, University of Cambridge, Madingley Road, Cambridge CB3 0HA}



\maketitle

\label{firstpage}

$$To\; appear\; in\; MNRAS$$

\begin{abstract}
Discoveries of exoplanets orbiting evolved stars motivate critical examinations of the dynamics of $N$-body systems with mass loss.  Multi-planet 
evolved systems are particularly complex because of the mutual interactions between the planets.  Here, we study the underlying dynamical mechanisms which can incite planetary escape in two-planet post-main sequence systems.  Stellar mass loss alone is unlikely to be rapid and high enough to eject planets at typically-observed separations.  However, the combination of mass loss and planet-planet interactions can prompt a shift from stable  to chaotic regions of phase space. Consequently, when mass loss ceases, the unstable configuration may cause escape. By assuming a constant stellar mass loss rate, we utilize maps of dynamical stability to illustrate the distribution of regular and chaotic trajectories in phase space.  We show that chaos can drive the planets to undergo close encounters, leading to the ejection of one planet.  Stellar mass loss can trigger the transition of a planetary system from a stable to chaotic configuration, subsequently causing escape.  We find that mass loss non-adiabatically affects planet-planet interaction for the most massive progenitor stars which avoid the supernova stage.  For these cases, we present specific examples of planetary escape.
\end{abstract}


\section{Introduction}

The first confirmed extrasolar planets were found to orbit evolved stars \cite{wolfra1992,wolszczan1994} and the last 5 years has
seen a resurgence of interest in this topic due to new discoveries (e.g. \cite{siletal2007,geietal2009,chaetal2011}).  
The majority of complimentary theoretical analyses has focused on single-planet, single-star evolved systems
\cite{villiv2007,villiv2009,Veras11,musvil2012,spimad2012} or systems with a belt or disc of material 
\cite{bonwya2010,bonetal2011,debetal2012}.  These investigations crucially establish  
physical and analytical frameworks from which to explore more complex systems in greater depth.

Just a few studies have considered multi-planet post-main-sequence (MS) evolution. \cite{dunclis98} performed long-term integrations of our Solar system assuming a constant Solar mass loss rate.  \cite{debsig2002}
demonstrated that multiple equal-mass planets on coplanar, circular orbits which are marginally stable on
the MS can become destabilised during post-MS evolution, because of the expansion of the Hill-stability limit
due to mass loss.  \cite{veretal12} showed that even Hill-stable pairs of planets can eventually become 
unstable, illustrating that instability occurs more readily than previously thought, and that planets do
not need to be closely packed to become unstable many Gyr later.  \cite{por2012} considered the evolution of
two planets in cataclysmic variable systems, but did not focus on planetary instability.

Although these initial studies have now broached the topic of planet-planet scattering amidst mass loss,
the detailed nature of this instability has yet to be explored.  Here, we pursue this line of investigation.
We consider a planetary system with two planets in a resonant or nonresonant configuration that is
stable on the MS and is perturbed when mass loss takes place.  We will show that 
the perturbation pushes the system into a chaotic regime through which instability manifests itself 
after the mass loss ceases.

The consequences of this late time instability may include collisions within the system
or escape from the system.  We focus on the later possibility, due to its potential relevance to the
purportedly vast free-floating planet population in the Milky Way \cite{Sumi11}, which is thought to outnumber
the Galactic population of bound planets.  
The existence of such substellar objects, known as free-floating or orphan planets (e.g. \cite{Lucas00,Zapat00,deletal2012}), may help us understand
the low-mass end of the initial mass function.  

The source of free-floating planets is unknown.  \cite{Veras12} considered the rate of ejections
in planet-planet scattering in MS systems, and found that even in the most optimistic case,
the rate is insufficient to generate the observed poputation.  Alternatively, \cite{Varvoglis} 
illustrated how a free-floating planet passing a stellar system could trigger dynamical instability
and incite planetary escape, leading to another free-floating planet.  \cite{zaktre2004} 
and \cite{vermoe2012} considered similar mechanisms for stellar flybys and star-planet flybys
in the Galactic disc, respectively.  These studies do not reproduce the required escape rate,
and potentially suggest that the free-floating planet population is a relic of the initial formation process.
The turbulent early-age birth environment of planets might represent the primary source,
as well as the frequent and slow flybys characteristic of young clusters 
\cite{adaetal2006,freetal2006,spuetal2009,maletal2011,boletal2012,parqua2012}.  Our focus
here, however, is on planetary escape in evolved systems, and on bounding the phase space in which this can occur.

The paper is organized as follows. In Section 2 we present some analytic and numerical 
results for the orbital element evolution of planets under stellar mass loss without taking into account the mutual planetary interaction. 
In Section 3 we present results that show the distribution of chaos and order in a system 
consisting of two massive planets and a star with constant mass; we discuss the resulting chaotic evolution and the destabilization 
of the system.  In Section 4 we present our results, which represent numerical simulations 
that combine both planetary interactions and stellar mass loss, and demonstrate the possible 
system destabilization and planetary escape. We present our main conclusions in Section 5.

\section{The two body problem of variable mass}
In the classical two body problem, the two bodies describe similar elliptic orbits around their center of mass when the motion is bounded. The motion is planar and the orbital elements of the two bodies (or the orbital elements of the relative motion of one body around the other) are constant. If, however, one of the bodies loses mass isotropically in all directions, then the evolution of the system is quite different. The orbital elements of the relative motion are no longer constant, and the evolution of the system depends on the mass loss function $m=m(t)$, where $m(t)$ is the sum of the masses of the two bodies. In this sense, the assumption of \cite{debsig2002} that mass loss by the star and mass gain by the planet are equivalent does not hold. A particular case, which we shall consider in this work, is a planetary system with a star and a planet with much smaller mass, where the star loses mass isotropically.  

The osculating elements of the relative motion are given by the differential equations \cite{Hadjid63}: 

\begin{equation} \label{DiffOEt}
\begin{array}{l}
\frac{da}{dt}=-a \frac{1+2e \cos f +e^2}{1-e^2} \frac{\dot m}{m}\\[0.2cm]
\frac{de}{dt}=-(e+\cos f)\frac{\dot m}{m} \\[0.2cm]
\frac{d\omega}{dt}=-\frac{\sin f}{e}\frac{\dot m}{m} \\[0.2cm]
\frac{df}{dt}=\frac{(G m)^{1/2}(1+e\cos f)^2}{[a (1-e^2)]^{3/2}} + \frac{\sin f}{e}\frac{\dot m}{m},  
\end{array}
\end{equation}      
where $a$ is the semimajor axis of the relative motion, $e$ the eccentricity, $\omega$ the argument of pericenter and $f$ the true anomaly.
One can readily check that the above equations admit the integral
\begin{equation} \label{Integral}
c^2=G m a(1-e^2),
\end{equation}
which in fact is the angular momentum integral.  We set $G = 1$ in the remainder of the paper.

In the following, we assume that the loss of the mass of the star is described by the Eddington-Jean's law
\begin{equation} \label{EJlaw}
\dot{m}=-\alpha m^n,
\end{equation}
where $\alpha$ is, in general, a small positive constant. The exponent $n$ defines the particular law, e.g. for $n=0$ we get a linear variation of mass, $m=m_0-\alpha t$, and for $n=1$ the exponential one, $m=m_0 e^{-\alpha t}$ . Also, we assume that the star loses mass down to a minimum limit $m_{min}=(1-\beta)m_0$, where the constant $\beta$ defines the final mass loss ratio ($0\leq \beta \leq 1$). This minimum mass is obtained at a time $t=t_\ell$, which is found through the solution of (\ref{EJlaw}) and depends on $m_0=m(0)$ and the parameters $\alpha$, $\beta$ and $n$.  

We introduce the normalized semimajor axis $\bar{a}=a/a_0$ and the normalized mass $\bar{m}=m/m_0$. The evolution of the osculating elements in the interval $0\leq t \leq t_\ell$ can be given as a function of the normalized mass $\bar{m}$, which decreases monotonically in time. Then equations (\ref{DiffOEt}) are written as
\begin{equation} 
\begin{array}{l}
\bar{m}\frac{d\bar{a}}{d\bar{m}}=-\frac{1+2e \cos f +e^2}{1-e^2} \bar{a} \\[0.2cm]
\bar{m}\frac{de}{d\bar{m}}=-(e+\cos f) \\[0.2cm]
\bar{m}\frac{d\omega}{d\bar{m}}=-\frac{\sin f}{e}\\[0.2cm]
\bar{m}\frac{df}{d\bar{m}}=-\frac{\bar{m}^{3-n}}{\epsilon} (1+e\cos f)^2 + \frac{\sin f}{e},  
\end{array}
\label{DiffOEnorm}
\end{equation}    
where
\begin{equation}\label{Epsilon}
\epsilon=\alpha c^3 m_0^{n-3}.
\end{equation}
From the above equations, and taking into account that $\bar{a}(0)=1$ and $\bar{m}(0)=1$ always, we conclude that two orbits with the same initial eccentricity $e_0$ and true anomaly $f_0$, show an equivalent evolution if the system is characterized by the same value of the parameter $\epsilon$. The initial value of the argument of pericenter $\omega_0$ affects only the evolution of $\omega=\omega(t)$.

In the above analysis, we considered a star and only one planet. If there are two, or more, planets, then the evolution of the osculating elements becomes more complicated, because the gravitational interaction between the planets affects in an important way the orbital evolution, as we shall see in the following sections. Only in the very special case where the masses of the planets could be considered as negligible would the evolution of each planet be determined by the system (\ref{DiffOEt}) or (\ref{DiffOEnorm}).

\subsection{Slow variation of mass - Evolution of the semimajor axis}
It can be proved \cite{Deprit83} that for a slow variation of mass and small eccentricities, the action $J=\sqrt{ma}$ is an adiabatic invariant, which implies that
\begin{equation} \label{adiabatic}
m(t)a(t)\approx constant.
\end{equation}
Thus, from the angular momentum integral (\ref{Integral}), we find that the eccentricity $e$ remains constant. From the above we see that the secular variation of the semimajor axis is given by
\begin{equation} \label{SemiAxisEvol}
a(t)\approx \frac{m_0}{m(t)} a_0,
\end{equation}
which implies that the semimajor axis increases monotonically, since the mass decreases. 
Particularly in the case of linear mass loss, $m=m_0-\alpha t$, we obtain the adiabatic estimation $a_{adiab}=a_0(1-\alpha t/m_0)^{-1}$  
given by Veras et al. (2011), while at the minimum mass limit the semimajor axis takes the value 
$$
a_\ell=a_0/(1-\beta).
$$

\subsection{The evolution of eccentric anomaly}
The variation of the eccentric anomaly $E$ in time, when the star loses mass isotropically, is necessary for estimating the time evolution of the eccentricity and the argument of pericenter, as we will show in the following paragraphs.
The eccentric anomaly obeys the equation (Hadjidemetriou, 1966)
\begin{equation} \label{EADeq}
\frac{dE}{dt}=\frac{1-e\cos E}{(1-e^2)^{1/2}} \frac{df}{dt}-\frac{\sin E}{1-e^2} \frac{de}{dt},
\end{equation}
where the derivatives of the true anomaly and the eccentricity are given by equations (\ref{DiffOEt}).  
In order to find the secular change of $E$, we ignore in (\ref{EADeq}) all the periodic terms and then, up to first order terms in eccentricity, we obtain 
\begin{equation} \label{EADeqEst1}
\frac{dE}{dt}=\frac{m^2}{c^3},
\end{equation}
or, by considering the normalization used in equations (\ref{DiffOEnorm}),
\begin{equation} \label{EADeqEst2}
\bar{m}\frac{dE}{d\bar{m}}=-\frac{\bar{m}^{3-n}}{\epsilon},
\end{equation} 
From the above equations we obtain that $dE/dt>0$ always holds true. The evolution of $E$ as a function of $\bar{m}$ depends essentially only on the system's parameters $\epsilon$ and $n$.  For the case of linear mass loss ($n=0$) we obtain
$$
E(\bar{m})=E_0\frac{1-\bar{m}^3}{3\epsilon},
$$
where $E_0=E(0)$, or
\begin{equation}\label{EtLinear}
E(t)=E_0+\frac{t}{3c^3} \big(3 m_0(m_0-at)+a^2t^2 \big). 
\end{equation} 
 
\subsection{Analytic estimate of the evolution of the eccentricity and the argument of pericenter}
Hadjidemetriou (1966) has given a series solution of equations (\ref{DiffOEt}) for the eccentricity and the argument of pericenter as a function of the eccentric anomaly $E$,
$$e=e_0+\sum_{j=1}{\epsilon^j f_j(E;e_0)} $$
$$\omega=\omega_0+\sum_{j=1}{\epsilon^j g_j(E;e_0)}, $$
which converge for $\frac{\epsilon}{e(1-e^2)^{3/2}}<1$. Assuming $\epsilon$ as a small parameter we can write, for small eccentricities, an approximate solution for $e=e(E)$ up to $O(\epsilon^3)$ as 
\begin{eqnarray}\label{SolEcc}
e&\approx& e_0+\epsilon \sin E + \frac{\epsilon^2}{2e_0} \sin^2E + \epsilon^2 E (3-n) \sin E 
\nonumber
\\
&+&  
\epsilon^3 E^2 (3-n)^2 \sin E +O(\epsilon^{i+1} E^i),\quad i\geq 3.
\end{eqnarray}   
The evolution of the eccentricity in time is given by considering the particular solution $E=E(t)$ of Eq. (\ref{EADeqEst1}).
Apart from the constant term $e_0$, the first two terms in (\ref{SolEcc}) are periodic terms of period $2\pi$ (one planetary revolution). The last two terms indicate also the same periodicity but with a secular variation of the amplitude of the periodic oscillations. Thus, an estimate of the maximum value reached by the eccentricity along the evolution can be approximated by the formula
\begin{equation}\label{MaxEstEcc}
A_e(E)=A_0+\epsilon^2|3-n| E + \epsilon^3(3-n)^2 E^2+O(\epsilon^{i+1} E^i),\quad i\geq 3.
\end{equation}
\begin{figure}[tb]
\centering
\includegraphics[width=8cm]{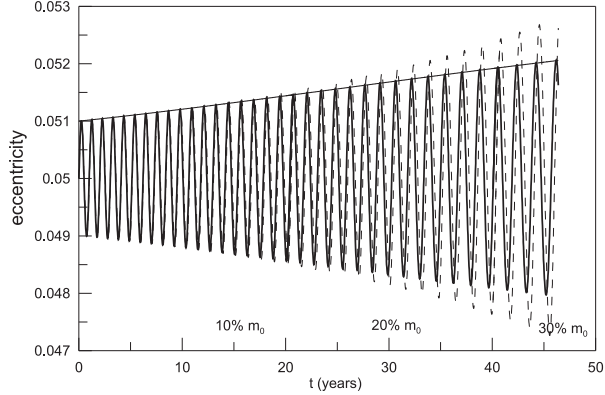} 
\caption{Evolution of eccentricity for the initial conditions $a_0=1$, $e_0=0.05$, $f_0=0$ and linear mass loss with $m_0=1$ and $\alpha=0.001$. The dashed line corresponds to the numerical solution of equations (\ref{DiffOEt}). The solid bold line is the analytic solution (\ref{SolEcc}), where $E$ have been substituted by eq. (\ref{EtLinear}). The estimate (\ref{MaxEstEcc}) of the maximum eccentricity is also shown by the thin solid line.}
\label{Figeccevol}
\end{figure}

The remaining terms $O(\epsilon^{i+1} E^i)$ are small for the first planetary revolutions, but they become important as the number of revolutions increases. In Fig. (\ref {Figeccevol}) we show an example of the evolution of eccentricity. We see that the analytic solution (\ref{SolEcc}) coincides very well with the real (numerical) solution but only for a relatively short time interval. 

Working in the same way as with eccentricity, we can write an approximate solution for the argument of pericenter
\begin{eqnarray}\label{SolOmega}
\omega&\approx& \omega_0+\epsilon \cos E + \epsilon^2 \frac{n-3}{e_0} E\cos E - \epsilon^3 E^2 \frac{(n-3)^2}{e_0} \cos E 
\nonumber
\\
&+& w_a + O(\epsilon^{i+1} E^i),\quad i\geq 3),
\end{eqnarray}
where 
$$
w_a=\epsilon^2 \frac{n}{2}E-\epsilon^3\frac{n(n-3)}{2}E^2
$$
is a secular component of the evolution. Note that $w_a$ is zero in the case of a linear mass loss ($n=0$).


\begin{figure}[tb]
$$
\begin{array}{cc}
\includegraphics[width=8cm]{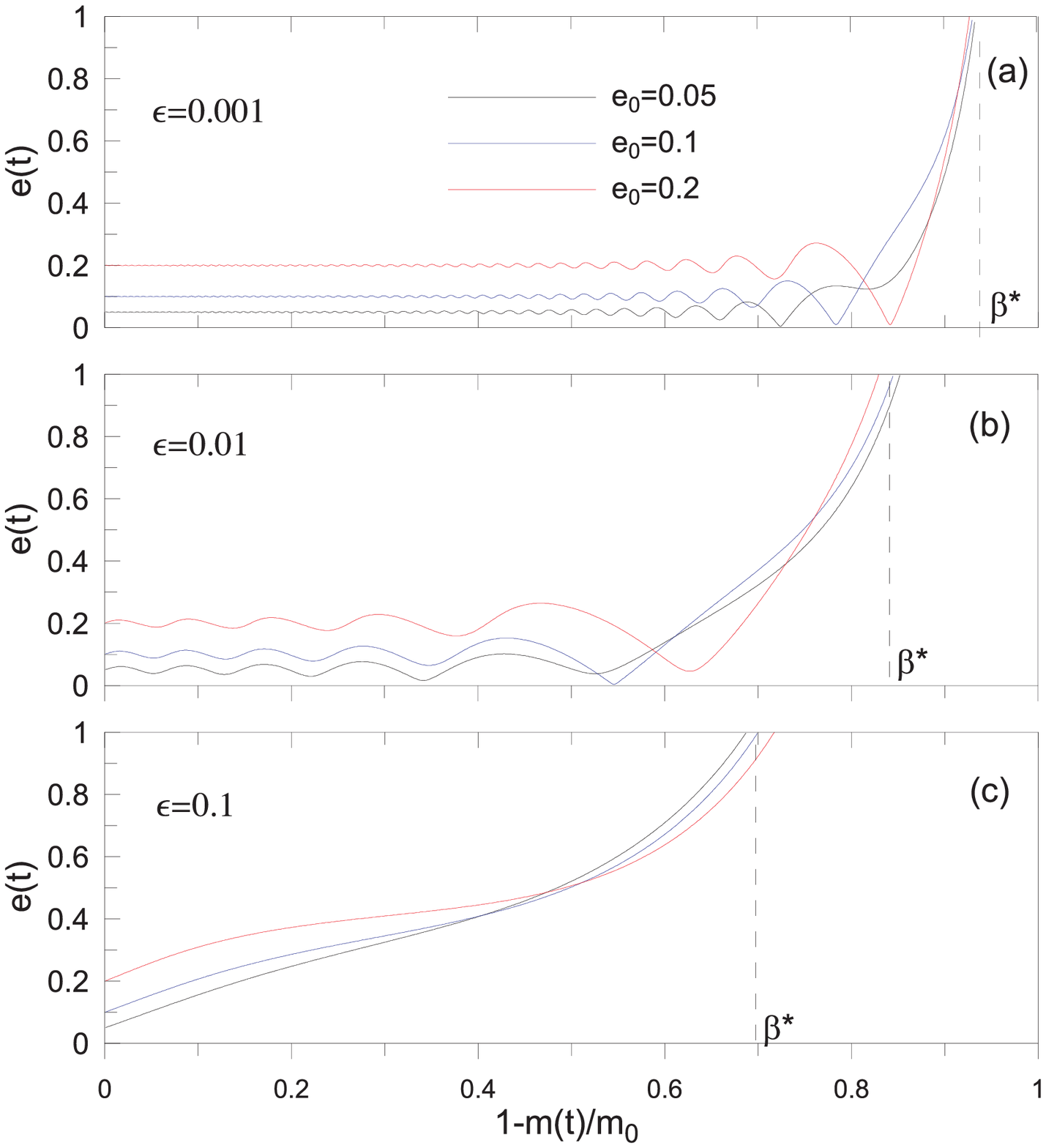} & \includegraphics[width=8cm]{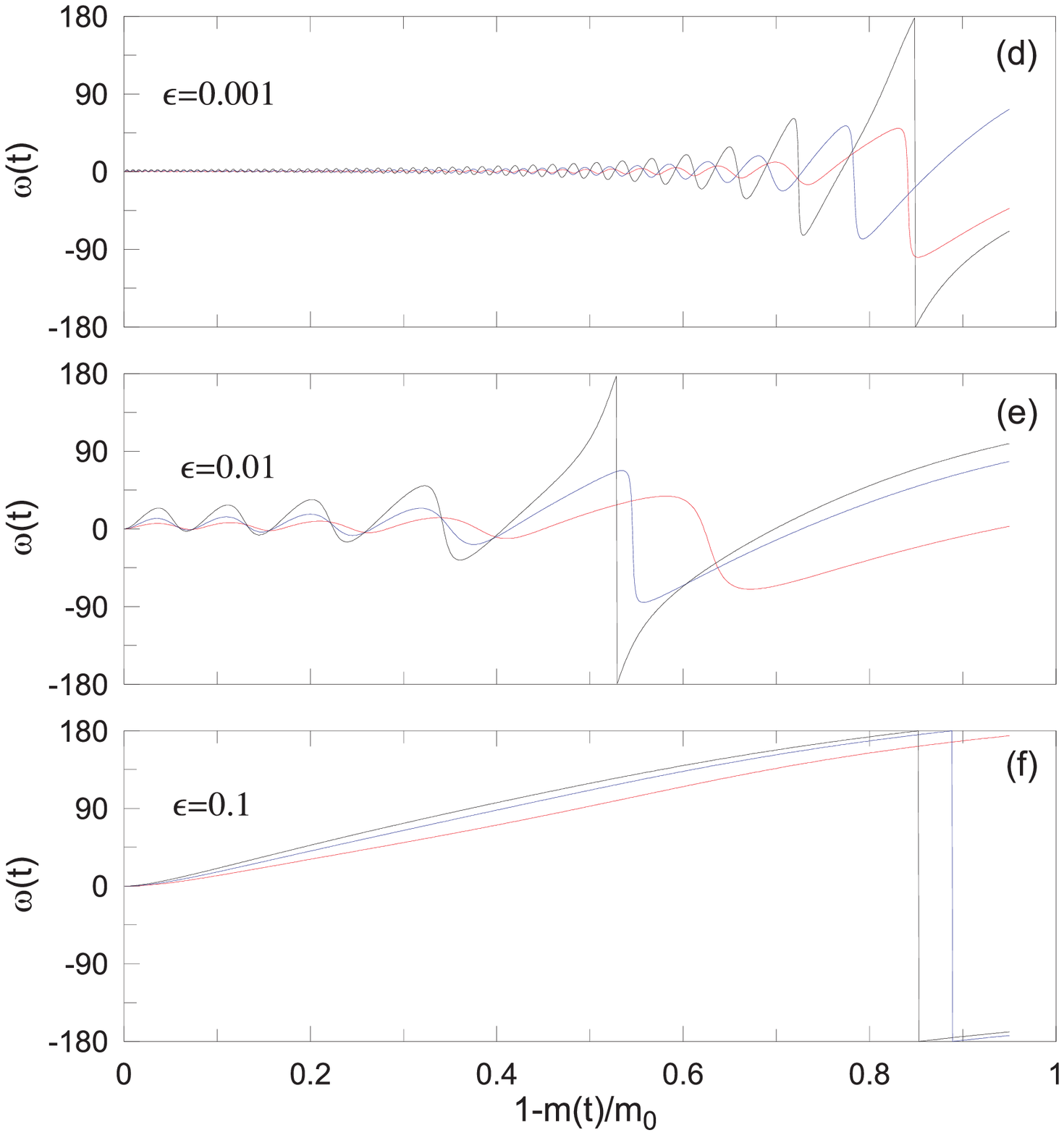} \\
(a) & (b)
\end{array}
$$
\caption{The evolution of the eccentricity (left panels) and the argument of pericenter (right panels) as the star loses mass with a linear rate, $m(t)=m_0-\alpha t$. Different color lines correspond to different initial values $e_0$ indicated in panel (a).}
\label{FigEccLinEvol} 
\end{figure}

\begin{figure}[tb]
\centering
\includegraphics[width=14cm]{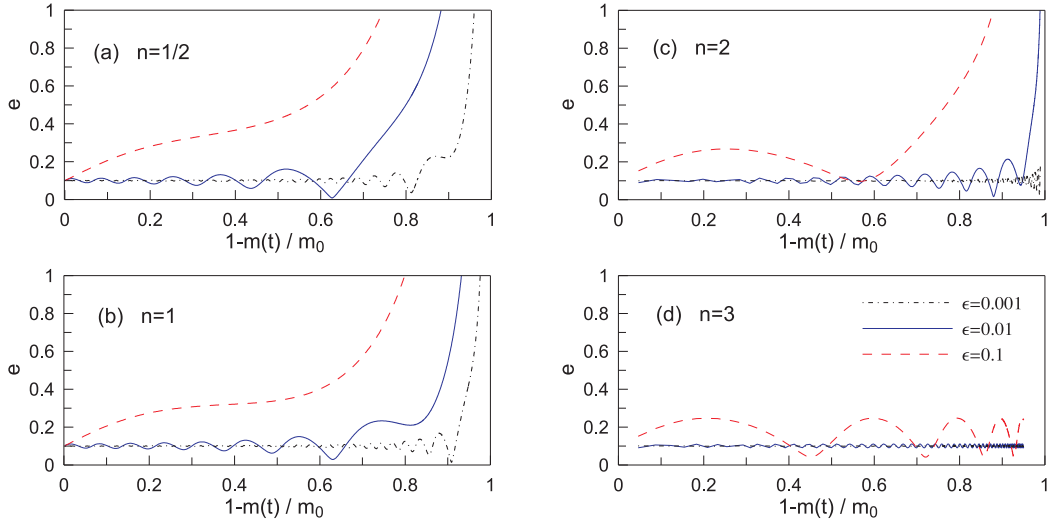} 
\caption{The evolution of the eccentricity for different exponents $n$ in the Eddington-Jean's law given by Eq. (\ref{EJlaw}). For each case we give the evolution for $\epsilon=0.001$, 0.01 and 0.1, as it is indicated in panel (d). The initial conditions are $e_0=0.1$, $f_0=\omega_0=0$.}
\label{Fignnneccevol}
\end{figure}

\subsection{Numerically calculated orbital evolution for one planet} \label{secUnperturbedNumOE}
The analytic estimates found above are sufficiently accurate only for time intervals where the mass loss does not exceed 20-30\% of the total mass. However, the numerical solution show that the eccentricity $e$ and the argument of pericenter $\omega$ continue to show oscillations with increasing amplitude for still larger mass loss, until their evolution enters a ``runaway regime'' where the eccentricity will eventually increase monotonically. The passage of the evolution to the ``runaway regime'' has been studied in detail by Veras et al. (2011) for the case of a linear mass loss. In the following we present some results obtained by numerical integrations. 

In Fig. \ref{FigEccLinEvol} (panels a,b and c) we show some typical examples of the evolution of the eccentricity $e$, where the horizontal axis represents the mass loss ratio $\beta=1-\bar{m}$, which can be mapped to the time $t$ through the particular mass loss law of eq. (\ref{EJlaw}). Here we consider the linear law ($n=0$). We observe that when the orbits enter the ``runaway regime'', we soon get $e>1$ and the planet escapes. Such an escape is obtained when the mass loss becomes sufficiently large, equal to a critical value $\beta^*$. This value depends generally on the initial conditions of the system, but when we start with small eccentricity values $e_0$, $\beta^*$ depends mainly on the parameter $\epsilon$. We observe that for $e=0.1$ the orbit enters the runaway regime from the beginning.  In the right panels d, e and f of Fig. \ref{FigEccLinEvol} we present the evolution of the argument of pericenter $\omega$ of the same orbits as in the left panels. Again we find that for small values of $\epsilon$, $\omega$ librates with an increasing amplitude, as it is indicated by the approximation of eq. (\ref{SolOmega}). However, in the case of $\epsilon=0.1$ (panel d), $\omega$ increases monotonically. In all cases $\omega$ completes one revolution at most.  

By considering $n\neq 0$, i.e. non-linear mass loss, we find that the evolution of the orbital elements is qualitatively similar to the linear case even though the runaway regime appears in general for larger values of mass loss. Some examples of the evolution of the eccentricity are shown in Fig. \ref{Fignnneccevol} for the cases $n=1/2$, $n=1$ (exponential decay), $n=2$ and $n=3$. For the case $n=2$ (panel c) and for $\epsilon=0.001$, the runaway behaviour appears only when the star mass becomes very small. An exception is the case $n=3$ (panel d) where the eccentricity continues to show small oscillations as $m\rightarrow 0$ and the runaway regime does not exist (at least for any $\epsilon<0.1$). From the series solution Eq. (\ref{SolEcc}) we obtain that for $n=3$ the amplitude of the oscillations is constant and the eccentricity is a periodic function of the eccentric anomaly (see also Hadjidemetriou, 1966). 

\begin{figure}[tb]
\centering
$$
\begin{array}{cc}
\includegraphics[width=7cm]{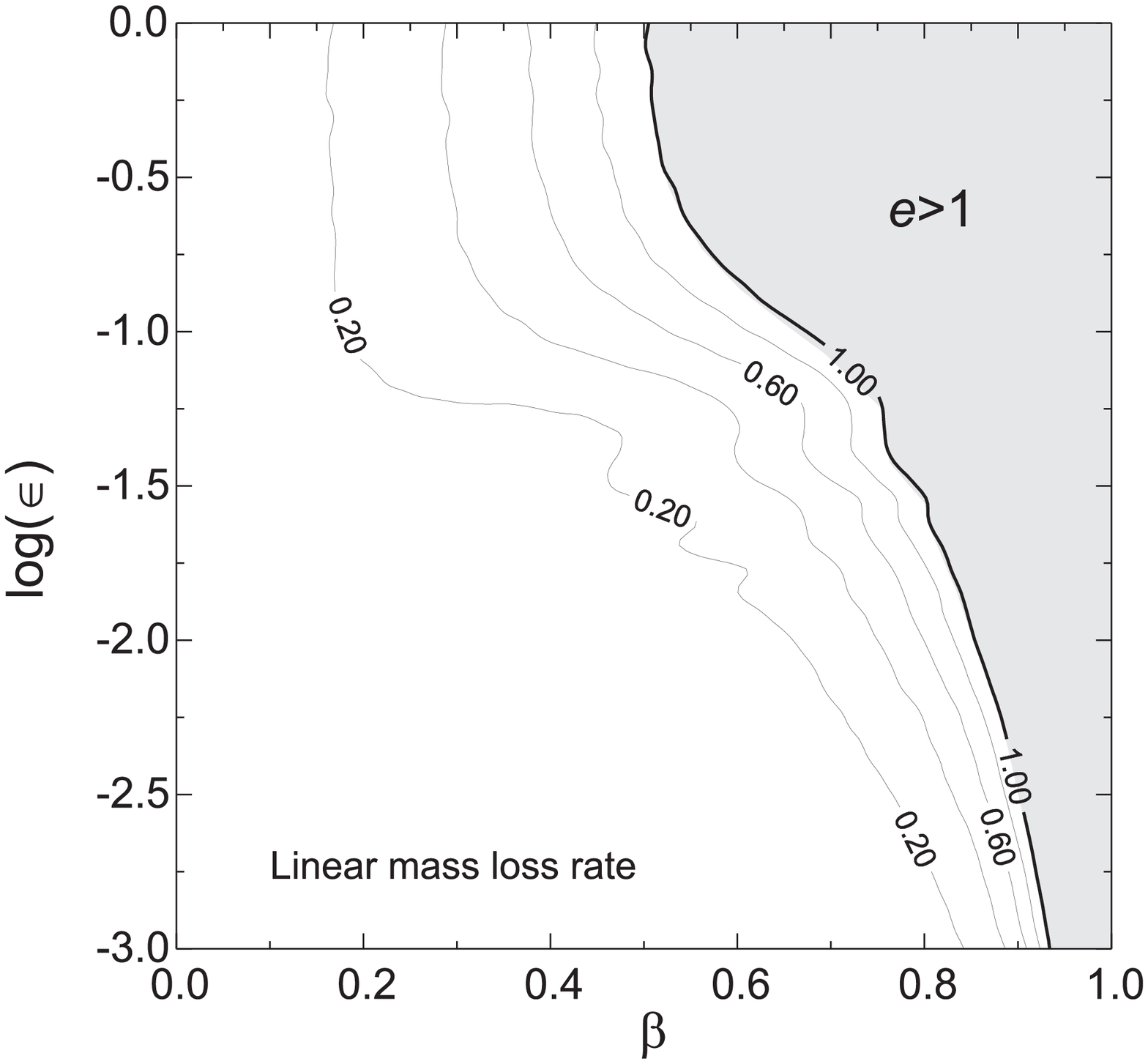} & \includegraphics[width=7cm]{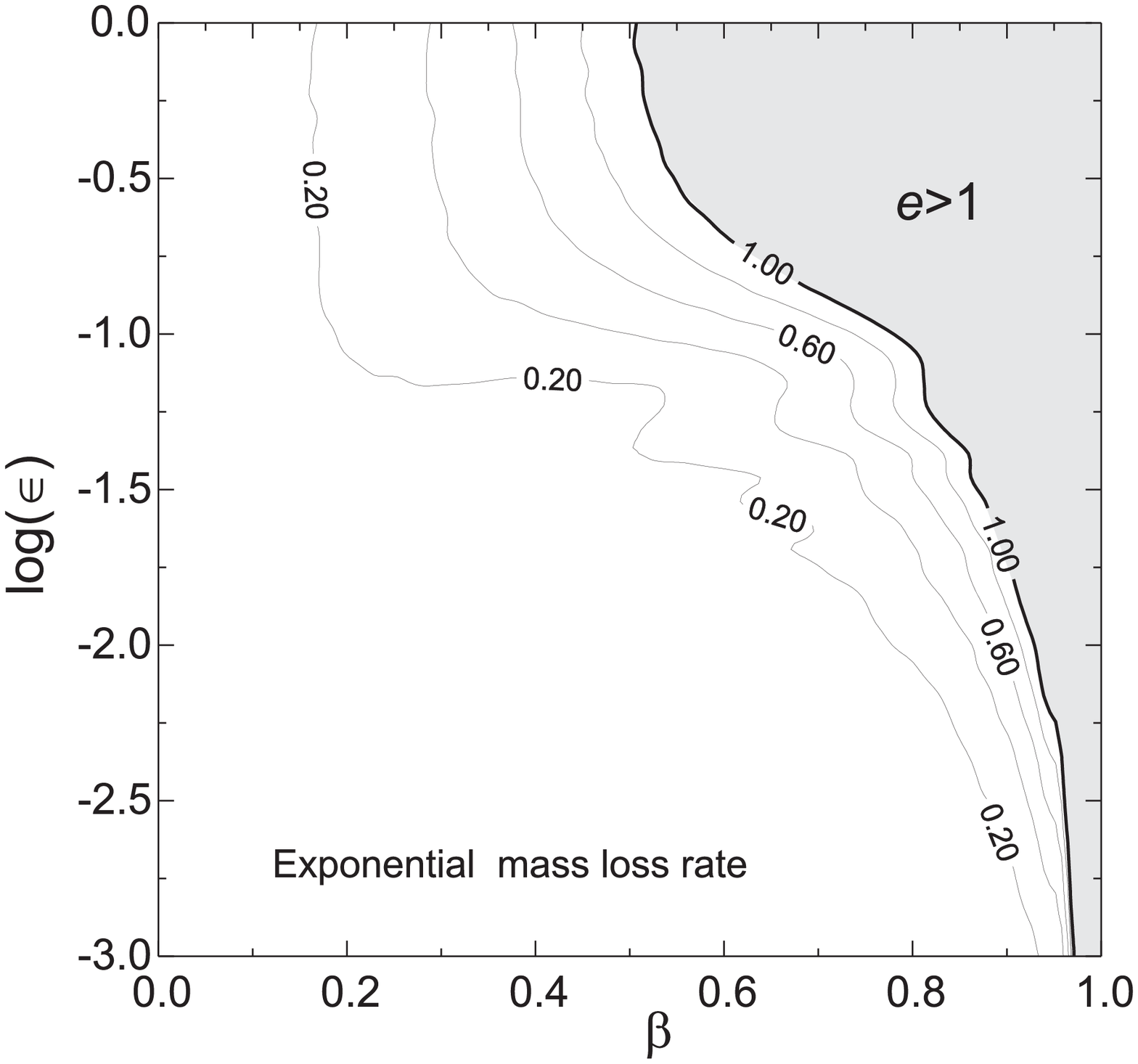} \\
(a) & (b)
\end{array}
$$
\caption{The mean eccentricity values obtained when the star mass loss ceases at a value $m=m_0(1-\beta)$ and for $\epsilon$ in the range $(0.001,1.0)$. The mean eccentricity is obtained after running 100 orbits with different initial conditions, $0<e_0\leq0.1$, $0\leq f_0\leq 2\pi$ and $\omega_0=0$. The gray region indicates the escape regime, where $e>1$. Left and right panels correspond to the linear ($n=0$) and the exponential ($n=1$) rate of mass loss, respectively.} 
\label{FigEccContours}
\end{figure}

So far we have shown that the planetary eccentricity may take large values either due to its large amplitude oscillations or due to the entrance in the runaway regime. Let us assume that the star mass loss starts at $t=0$, when $e=e_0$, and  continues up to a particular mass loss ratio value $\beta$, which is reached after time $t=t_\ell$. Then, at the end of the star mass loss event, the planetary orbit has reached a new eccentricity value $e_\ell=e(t_\ell)$. In the contour maps of Fig. \ref{FigEccContours} we present the value of $e_\ell$ in the parameter domain $0\leq\beta<1$ and $0.001\leq\epsilon<1$. The cases of linear (left panel) and exponential (right panel) rates of mass loss are given. The tabulated eccentricity value is the average eccentricity obtained at $t=t_\ell$ from 100 orbits with initial eccentricity and true anomaly values that were randomly selected in the intervals $(0,0.1)$ and $(0,2\pi)$, respectively. We see that for small values of the parameter $\epsilon$ (e.g. $\epsilon\leq10^{-3}$), a significant increment of the eccentricity is obtained only after a significant mass loss (e.g. more than 80\% in the linear case and more than 90\% in the exponential case). As $\epsilon$ takes larger values, a significant eccentricity increment is observed for smaller amounts of lost mass. The escape regime ($e>1$) is indicated by the gray shaded region and occupies the region of large values of $\beta$ and $\epsilon$, as was expected. By computing the same maps for $n=2$ and $n=3$, we obtain similar plots.  However, these curves, which correspond to the same eccentricity levels, are obtained for larger $\beta$ values and yield smaller escape regimes.       
    
\subsection{Resonant evolution}
As previously mentioned, when there are at least two planets in the system, the evolution does not depend on the mass loss of the star only; the gravitational interaction between the planets plays an important role. In order to study this effect, and reveal the extent to which the gravitational interaction influences the evolution, we consider here a planetary system with two planets of negligible mass amidst mass loss.  In this case, no gravitational interaction exists between the planets.

Consider a system with two massless planets $P_i$, $i=1,2$, with semimajor axes $a_i$ and angular momenta $c_i$. The evolution of each planet depends on the value of the parameter $\epsilon$, namely $\epsilon_i=\alpha c_i^3 m_0^{n-3}$. Because $\epsilon_i$ is different for each planet, each planet then follows a different evolution. If $P_1$ is the outer planet ($a_2>a_1$), then we obtain $\epsilon_2>\epsilon_1$. The mean motion resonance is defined by the mean motion ratio $n_1/n_2$, where $n_i=m^{1/2} a_i^{-3/2}$, which is given by the relation
\begin{equation} \label{ResEvol}
\frac{n_1}{n_2}=\left ( \frac{c_2}{c_1} \right )^3 \left ( \frac{1-e_1^2}{1-e_2^2} \right )^{3/2}=\frac{\epsilon_2}{\epsilon_1} + O(e_i^2).     
\end{equation}
Thus two resonat planets with orbits of low eccentricities preserve their mean motion resonance $\frac{n_1}{n_2}\approx \frac{p+q}{p}$, $p,q$ integers, under the star mass loss and before entering the runaway regime, where the eccentricities take large values. Also, the preservation of the resonance and the libration of the argument of pericenter, mentioned in section \ref{secUnperturbedNumOE}, indicate that the resonant angles $\Delta\omega=\omega_2-\omega_1$ and $\theta_i=p \lambda_1-(p+q)\lambda_2+q \omega_i$, $i=1,2$, should also librate, until a significant stellar mass loss is reached.

\begin{figure}[htb]
\centering
$$
\begin{array}{cc}
\includegraphics[width=7cm]{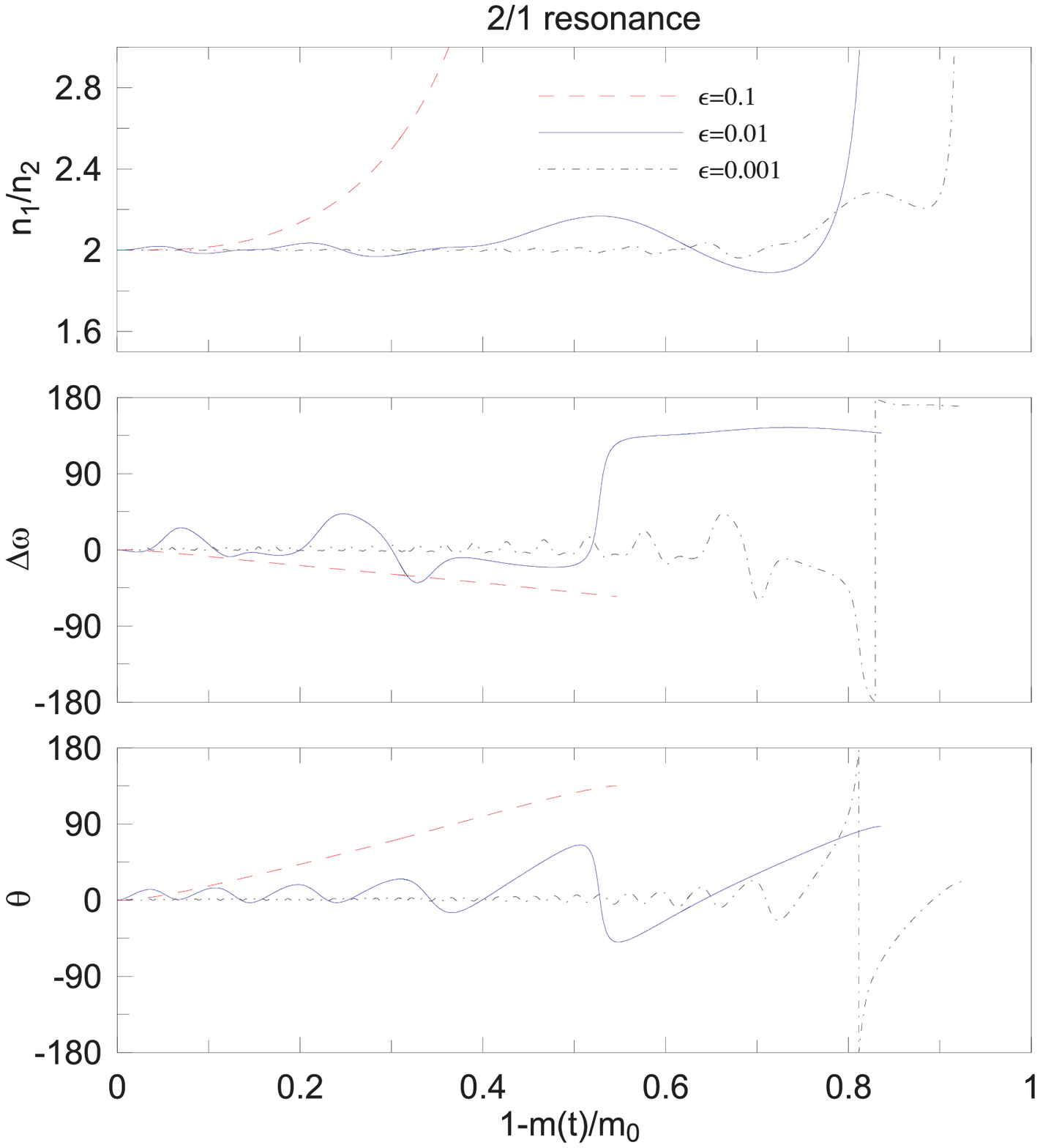} & \includegraphics[width=7cm]{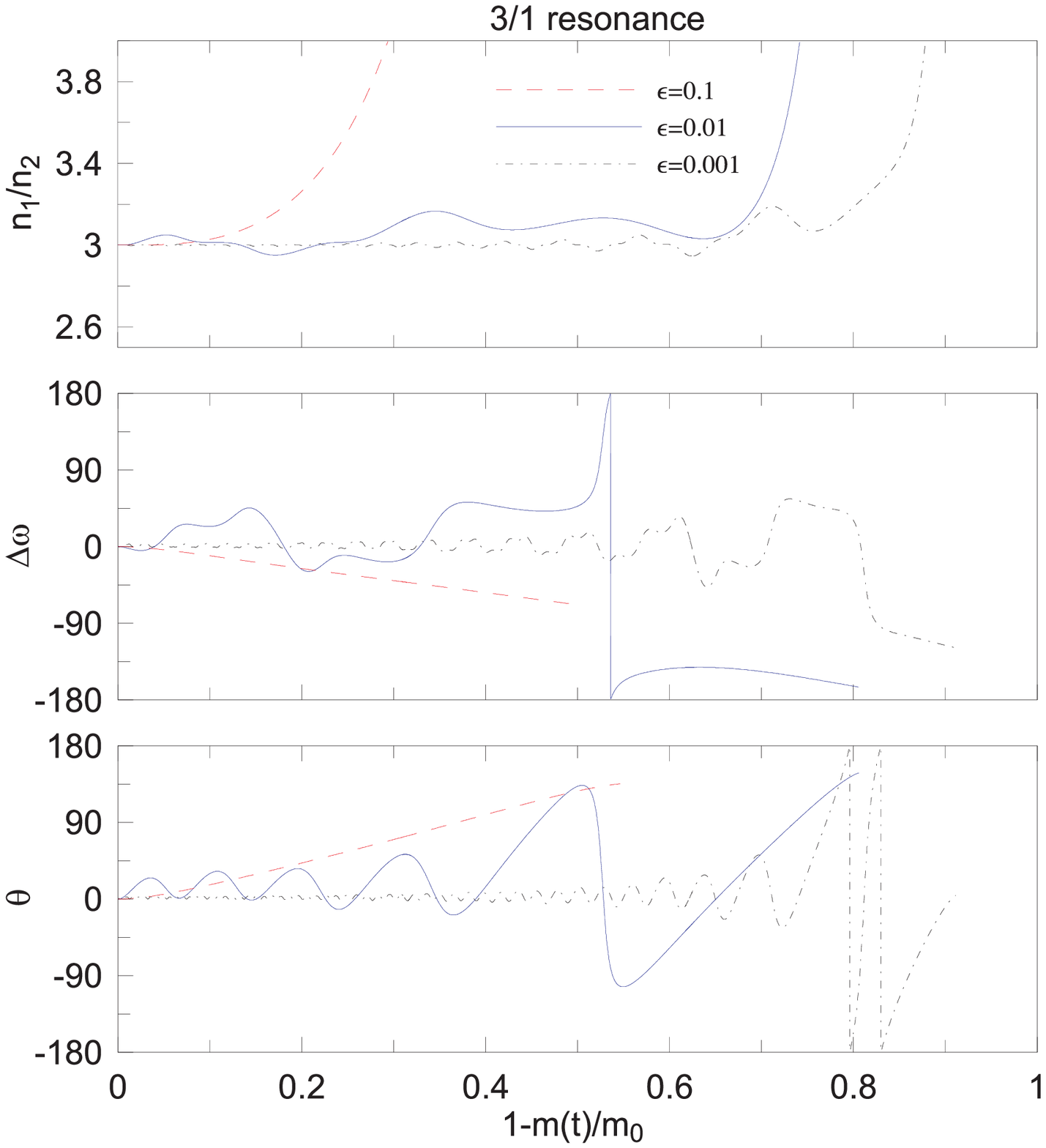} \\
(a) & (b)
\end{array}
$$
\caption{The evolution of the mean motion ratio and the resonant angles $\Delta \omega$ and $\theta=\theta_1$ as the star loses mass at a linear rate, $m(t)=m_0-\alpha t$. (a) starting from the 2/1 resonance and (b) starting from the 3/1 resonance. The initial conditions are $e_0=0.1$, $f_0=\omega_0=0$. } 
\label{FigResEvol}
\end{figure}

In Fig. \ref{FigResEvol} we present the evolution of the mean motion ratio and of the resonant angles for the 2/1 (left panel) and 3/1 (right panel) resonance and for $\epsilon_1=10^{-3}$, $10^{-2}$ and $10^{-1}$. For the outer planet it is $\epsilon_2\approx 2\epsilon_1$ and $\epsilon_2\approx 3\epsilon_1$ for the 2/1 and 3/1 resonance, respectively. For both resonances we see that the system evolves in a similar way. For small mass loss parameters $\epsilon_i$,  $n_1/n_2$ is almost constant before a significant amount of stellar mass loss. First the outer planet enters the runaway regime and then a rapid increment of $n_1/n_2$ is observed. As we consider larger values of $\epsilon$, the system leaves the resonance at a smaller mass loss ratio $\beta=1-\bar{m}$. As far as the system remains close to the resonance, the resonant angles show small oscillations around their initial value. Afterwards, the amplitude of the libration increases and the evolution of the resonant angles might end up in circulation. For large values of $\epsilon$ (e.g. for $\epsilon=0.1$), we obtain a slow monotonic variation of the angles. Particularly, $\Delta \omega$ decreases while $\theta_1$ increases. We stop the evolution when the outer planet escapes, i.e., when $e_2>1$, where the mean motion is no longer defined.   

\section{Chaos and ejection in two-planet systems}

Here we study the evolution of a two-planet system where the mutual planetary gravitational interaction is included but the star does not lose mass. Thus we establish a dynamical basis with which to compare the mass loss case in the next section.

\subsection{Model and methods}
The study of the dynamics of a system consisting of a star of mass $m_0$ and two planets $P_i$, $i=1,2$ of masses $m_i$ can be studied by using the model of the general three body problem (GTBP)  
\begin{equation} \label{GTBPdeqs}
\mathbf{\ddot{r}}_i=G\sum_{j=0} \frac{m_j}{r_{ij}^3} \mathbf{\delta r}_{ij}, \qquad (i\neq j,\; i=1,2),
\end{equation}
where $\mathbf{r}_i$ indicates the position of the planet and $\mathbf{\delta r}_{ij}=\mathbf{r}_j-\mathbf{r}_i$. We restrict our study to the planar case, $\mathbf{r}_i=(x_i,y_i)$ where the inertial frame $Oxy$ is centered at the barycenter. Also 
we normalize the units by setting $G=1$ and $m_0+m_1+m_2=1$. In the following, we should take always the $P_1$ to be initially the inner planet and $P_2$ the outer one in the sense that $a_1<a_2$, where $a_i$ is the semimajor axis of $P_i$.

Although the system appears having four degrees of freedom, we can use the angular momentum integral and a rotating frame and reduce the system to three degrees of freedom (Hadjidemetriou, 1975). In such a system, regular planetary orbits correspond to quasiperiodic trajectories, which twist on invariant tori in phase space according to the KAM theorem. However the mutual planetary interaction destroys the integrability of the system and chaotic orbits coexist beside the regular ones. 

\begin{figure}[htb]
\centering
\includegraphics[width=5cm]{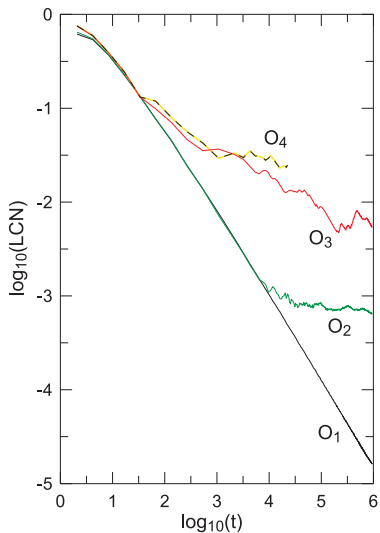} 
\caption{The LCN evolution for a regular orbit ($O_1$), chaotic orbits ($O_2$, $O_3$) and a chaotic orbit ($O_4$) that shows a close encounter after 3.5 Ky  (we note that 6.28 time units correspond to about 1 year). The initial conditions are $a_1=1$, $a_2=1.74$, $\omega_1=0$, $\omega_2=180^\circ$, $M_1=M_2=0$, $e_1=0.1$ and $e_2=0.05$, $0.1$, $0.25$ and $0.26$ for the orbits $O_1$-$O_4$, respectively.}
\label{FigLCNplot}
\end{figure}

Many numerical tools for the detection of chaos have been proposed. For example, in planetary dynamics the chaos indicators MEGNO, RLI and FLI have been used (Go\'zdziewski, 2005;  S\'andor et al, 2007; Voyatzis, 2008, respectively). In the present work we compute the maximal Lyapunov characteristic number (LCN), which is the classical measure for the average exponential divergence of nearby orbits.  An example of the evolution of LCN for some different trajectories of the GTBP is shown in Fig. \ref{FigLCNplot} for $10^6$ time units (or $\approx 150$ ky).  A regular evolution is described by an LCN evolution that tends to zero, as e.g. the orbit $O_1$. For chaotic orbits LCN tends to a positive value, as e.g. in the orbits $O_2$ and  $O_3$. This value is used to form {\em dynamical stability maps} (DS maps).  In these we define plane grids of initial conditions presented with a color scale representing the LCN value after a particular integration time $t_{max}$\footnote{An application of LCN maps of dynamical stability to planetary dynamics is given in  Hadjidemetriou and Voyatzis (2011a).}. For the numerical integration of Eq. (\ref{GTBPdeqs}) we use the Bulirsch-Stoer integrator. When planetary close encounters occur, the integration may break, in the sense that the integration step becomes very small in order for the method to preserve the requested integration accuracy (e.g. the orbit $O_4$). Such cases always correspond to strongly chaotic orbits and in the DS maps are presented by the value $LCN=1$ (light-coloured regions).

\begin{figure}[htb]
\centering
\includegraphics[width=7cm]{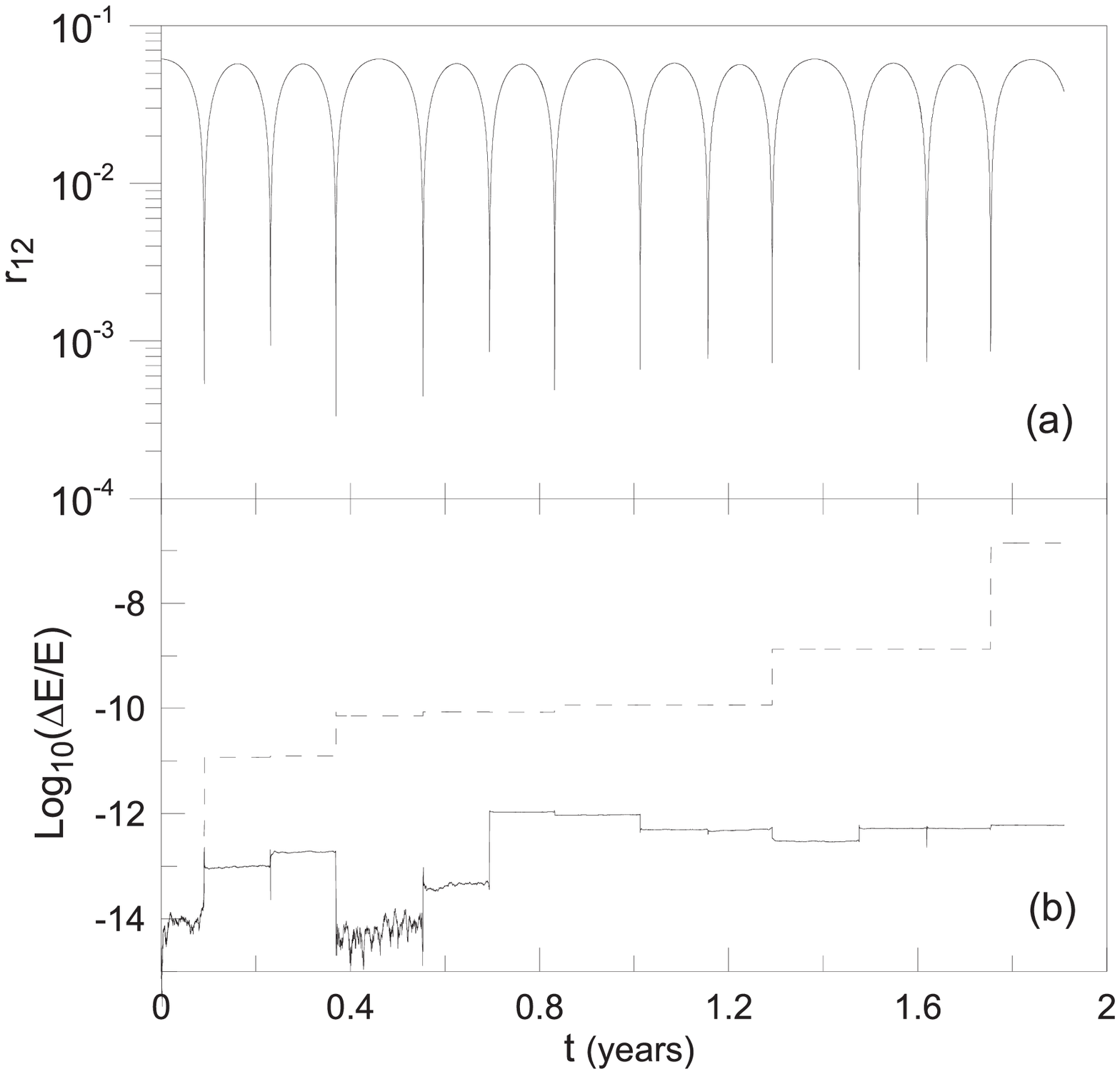} 
\caption{An example of a planetary evolution (1/1 resonant) that shows a sequence of close encounters  a) the planetary distance $r_{12}$ b) the evolution of the energy error along the numerical integration with and without regularization (solid and dashed curves, respectively). }
\label{FigAcc}
\end{figure}

In cases where we need to follow the evolution after a close encounter and without loss of accuracy, we apply a regularization of the collision singularities. In this paper we apply a Levi-Civita transformation when the gravitational interaction between any two bodies of the system becomes very strong (Marchal, 1990; Aarseth, 2003). In Fig. \ref{FigAcc} we present an example of evolution with a sequence of planetary close encounters. The two planets are placed inside the Hill radius and evolve in a satellite configuration. In panel (a) we present the planetary distance $r_{12}$. We integrate the system using two methods. In the first we integrate equations (\ref{GTBPdeqs}) using the BS integrator with accuracy $10^{-14}$. When the step becomes very small, instead of stopping the integration, we reduce the requested accuracy. In the second method we use the same integrator as above but we switch on the regularized equations during planetary close encounters. In panel (b) of Fig \ref{FigAcc} we  present the error in energy along the trajectory in both cases. We can observe that without regularization the error increases after each close encounter, while the integration of the regularized equations preserves the accuracy at a very good level.

\begin{figure*}
\includegraphics[width=12cm]{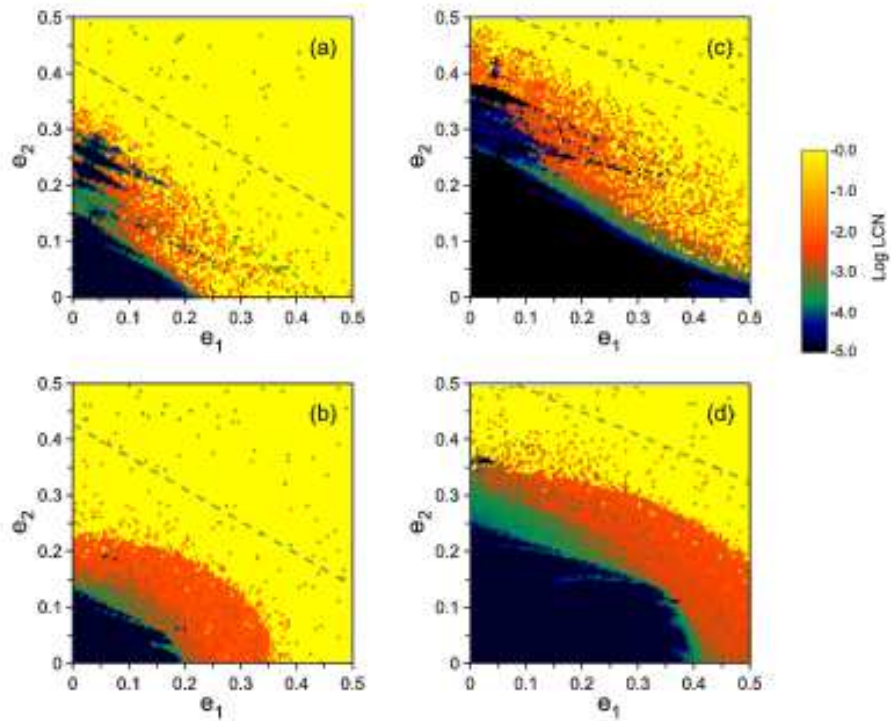} 
\caption{DS maps in the eccentricity plane for $n_1/n_2\approx 2.29$ ($a_1=1.0$, $a_2=1.74$) (panels a,b) and  
$n_1/n_2\approx 3.29$ ($a_1=1.0$, $a_2=2.21$) (panels c,d) . The top panels correspond to $m_1=m_J$, $m_2=0.2m_J$ and the bottom ones to $m_1=m_J$, $m_2=2m_J$. For all orbits we set as initial conditions $\omega_1=0^\circ$, $\omega_2=180^\circ$, $M_1=M_2=0^0$. The dashed curve indicates the collision line.}   
\label{FigNRmaps}
\end{figure*}

\subsection{DS maps} \label{secDSM}
We restrict our numerical simulations to planetary systems with the inner planet having mass $m_1=0.001$ (approximately Jupiter's mass, $m_J$) and semimajor axis $a_1=1$ AU. The outer planet is either lighter ($m_2=0.2m_J$) or heavier ($m_2=2m_J$) than the inner one and with semimajor axis in the range $1.5<a_2<2.5$ AU.   

\subsubsection{Non-resonant motion}
For non-resonant motion, the main source of chaos is planetary close encounters, which can happen when the initially elliptic orbits of the two planets intersect.  
In Fig. \ref{FigNRmaps} we present DS maps for non resonant motion. The maps are defined by grids of initial conditions in the plane of eccentricities.  The other orbital elements are fixed and are given in the caption. In panels (a) and (b) the orbits correspond to an initial mean motion ratio $n_1/n_2=2.29$. In case (a), where the outer planet ($P_2$) is more massive than the inner one ($P_1$), we observe that regular orbits exists only in a well-defined region at small eccentricity values. Some small stable regions appear for $e_2$ up to 0.3 but only for $e_1<0.1$. A similar picture is obtained also for the case (b), where the inner planet is lighter than the outer one. Now the stable orbits are confined in a region of small eccentricities. Above the line $a_1(1+e_1)=a_2(1-e_2)$ (collision line), which is represented by the dashed line in the dynamical maps, the planetary orbits intersect and strong chaos appears. We obtain the same dynamical picture in panels (c) and (d), where $n_1/n_2=3.29$. In this case the outer planet starts with a larger semimajor axis than in the previous case. The planetary interactions are weaker, the stable region covers a larger domain and the collision line appears at higher eccentricity values.

\begin{figure*}
\includegraphics[width=12cm]{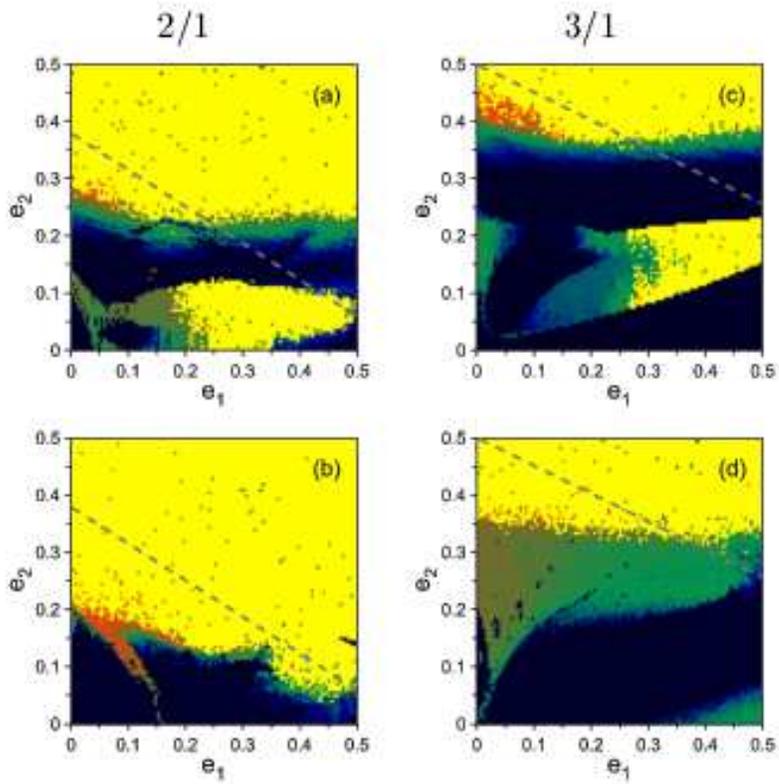} 
\caption{DS maps on the eccentricity plane for the 2/1 and 3/1 resonances. The top panels correspond to $m_1=m_J$, $m_2=0.2m_J$ and the bottom ones to $m_1=m_J$, $m_2=2m_J$. The initial semimajor axes are $a_1=1.0$ and $a_2=1.59$ and $2.08$ for the 2/1 and 3/1 resonances, respectively. For all orbits we start with $\omega_1=0^\circ$, $\omega_2=180^\circ$, $M_1=M_2=0^0$. The dashed line indicates the planetary collision line.} 
\label{FigRESmaps} 
\end{figure*}

\subsubsection{Resonant motion}
In Fig. \ref{FigRESmaps} we present DS maps for $n_1/n_2=2.0$ (panels a and b) and $n_1/n_2=3.0$ (panels c and d). In these cases we obtain a complex structure in phase space where regions of stable and unstable motion are intermingled (Michtchenko et al, 2008a,b; Hadjidemetriou and Voyatzis, 2009).  Since resonances may offer phase protection mechanisms, stable long term evolution might appear even for intersecting orbits.
We observe in the DS maps that we can have regular evolution when the orbit of the inner planet is very eccentric. However when the eccentricity of the outer planet becomes larger than about 0.2 we obtain a wide chaotic sea where trajectories are strongly chaotic. For the  3/1 resonant case with $m_1>m_2$ we observe a zone of regular orbits for $0.25<e_2<0.35$. The existence of such zones and, generally, the distribution of chaotic and regular regimes in phase space is significantly affected by the families of periodic orbits (symmetric or asymmetric) that, generally, exist in the resonant regions (see e.g. Voyatzis, 2008).   

\begin{figure*}
\includegraphics[width=14cm]{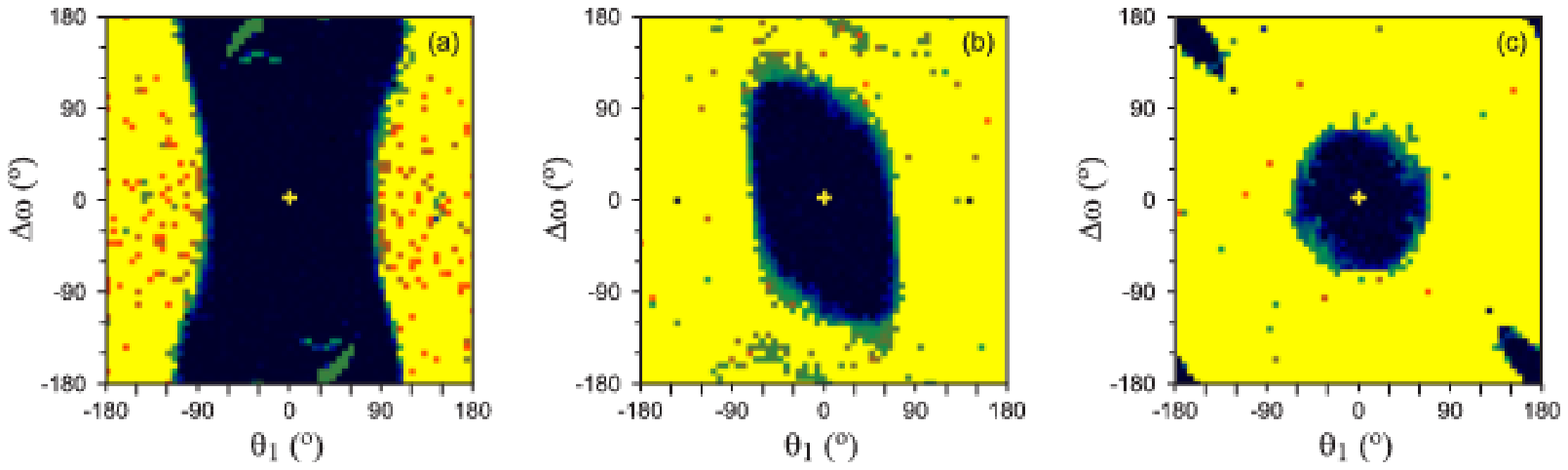} 
\caption{DS maps around a 2/1 resonant stable periodic orbit on the plane of resonant angles $(\theta_1, \Delta \omega)$. The periodic orbit is located at (0,0) with (approximate) initial conditions a) $a_1=1$, $a_2=1.582$, $e_1=0.3$, $e_2=0.063$  b) $a_1=1$, $a_2=1.580$, $e_1=0.5$, $e_2=0.166$ and c) $a_1=1$, $a_2=1.580$, $e_1=0.7$, $e_2=0.318$. For all cases it is $\omega_1=\omega_2=0$, $M_1=0$, $M_2=180^\circ$ and $m_1=m_J$, $m_2=2m_J$. The cross in the center indicates the position of periodic orbit.}  
 \label{FigPOmaps} 
\end{figure*}

\subsubsection{Exact resonances - periodic orbits}    
Resonant motion is related with the existence of monoparametric families of periodic orbits in phase space. Such periodic orbits are also called exact resonances (Beaug\'e et al, 2003), which can be stable or unstable and contribute significantly to the topology of phase space and the existence of order and chaos. In particular, stable periodic orbits are surrounded by regions of regular motion where the resonant angles $\Delta\omega$, $\theta_1$ and $\theta_2$ librate. Outside these regions stable motion can be also obtained, where some of the resonant angles circulate (Michtchenko et al, 2008a,b;  Voyatzis, 2008). In Fig. \ref{FigPOmaps} we present maps of dynamical stability which show the distribution of regular motion around 2/1 resonant periodic orbits, which belong to the symmetric family of periodic orbits for $m_1=m_J$ and $m_2=2m_J$ with configuration ($\theta_1,\Delta\omega)=(0,0)$ (family $S_1$ in Voyatzis et al, 2009).  The maps are defined on the plane of resonant angles $(\theta_1, \Delta\omega)$ and we see that the regular region, which surrounds the periodic orbit at $(0,0)$, shrinks as the eccentricities increase. 

Resonant families of stable periodic orbits are of particular importance, since they form paths in phase space for a migrating planetary system (Lee and Peale, 2002; Ferraz-Mello et al, 2003, Hadjidemetriou and Voyatzis, 2010, 2011b). So, regions around stable periodic orbits are candidates for containing a planetary system after a migrating process and a capture in resonance.

\subsection{Escape of planets from planetary systems in a chaotic region}\label{chaotic}
In this section we study the evolution of a planetary system under the gravitational interaction between the two planets, when the system is initially located in a chaotic region.  The mechanism that transfers an initially stable system to a chaotic region will be studied in the next section.

\begin{figure}[tb]
\centering
\includegraphics[width=9cm]{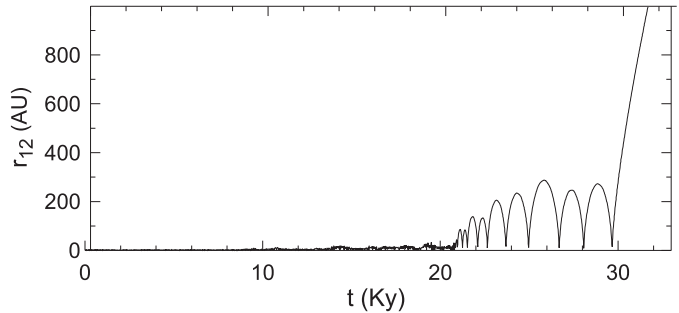} 
\caption{The mutual distance of planets $P_1$ and $P_2$ along a strongly chaotic orbit with initial conditions as those in the DS map of figure \ref{FigRESmaps}a and for $e_1=0.05$ and $e_2=0.3$. The planet $P_2$ escapes after about 30 Ky.}
\label{FigEvolRmin}
\end{figure}

\begin{figure}[tb] 
$$
\begin{array}{cc}
\includegraphics[width=6cm]{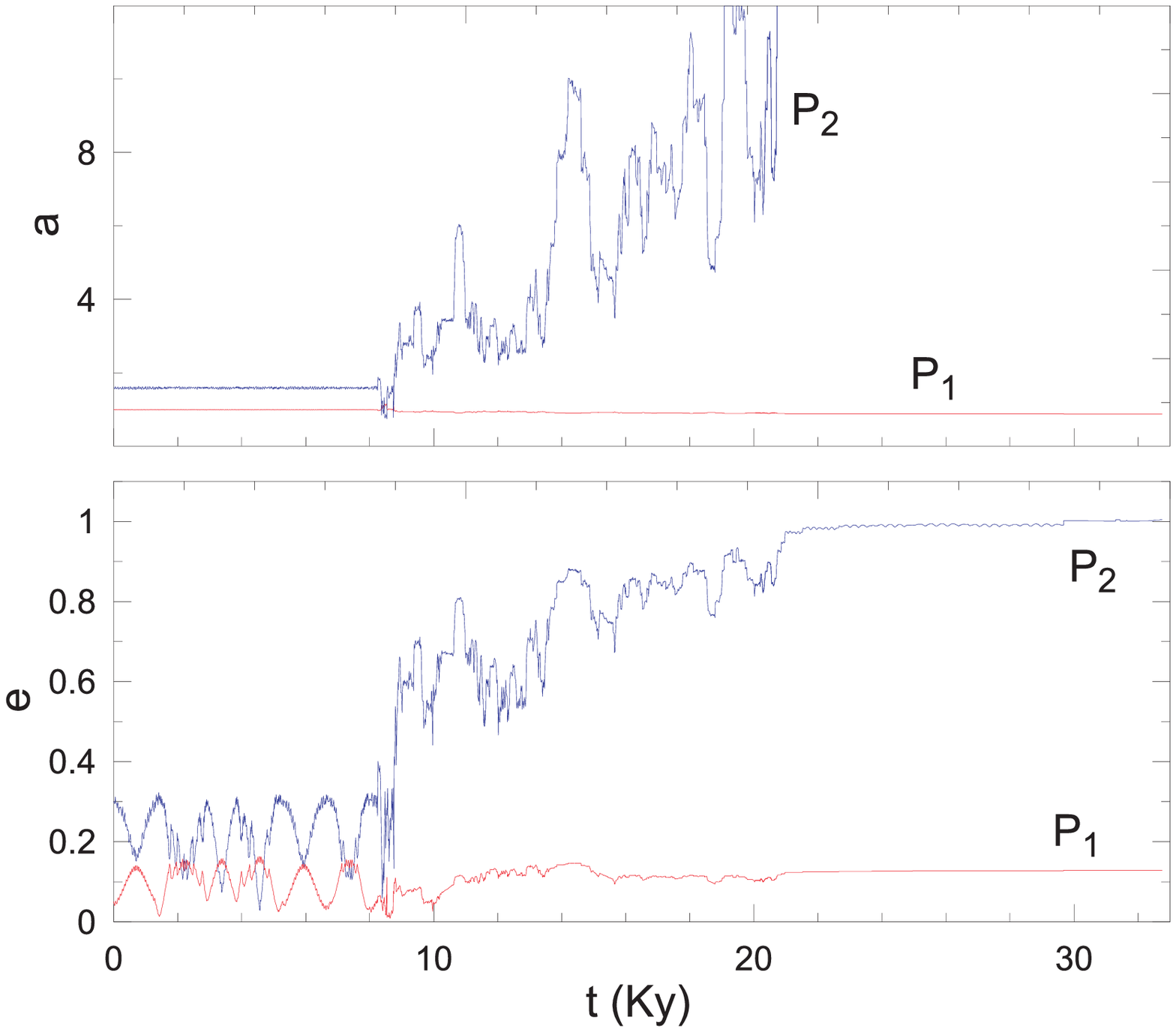} & \includegraphics[width=6cm]{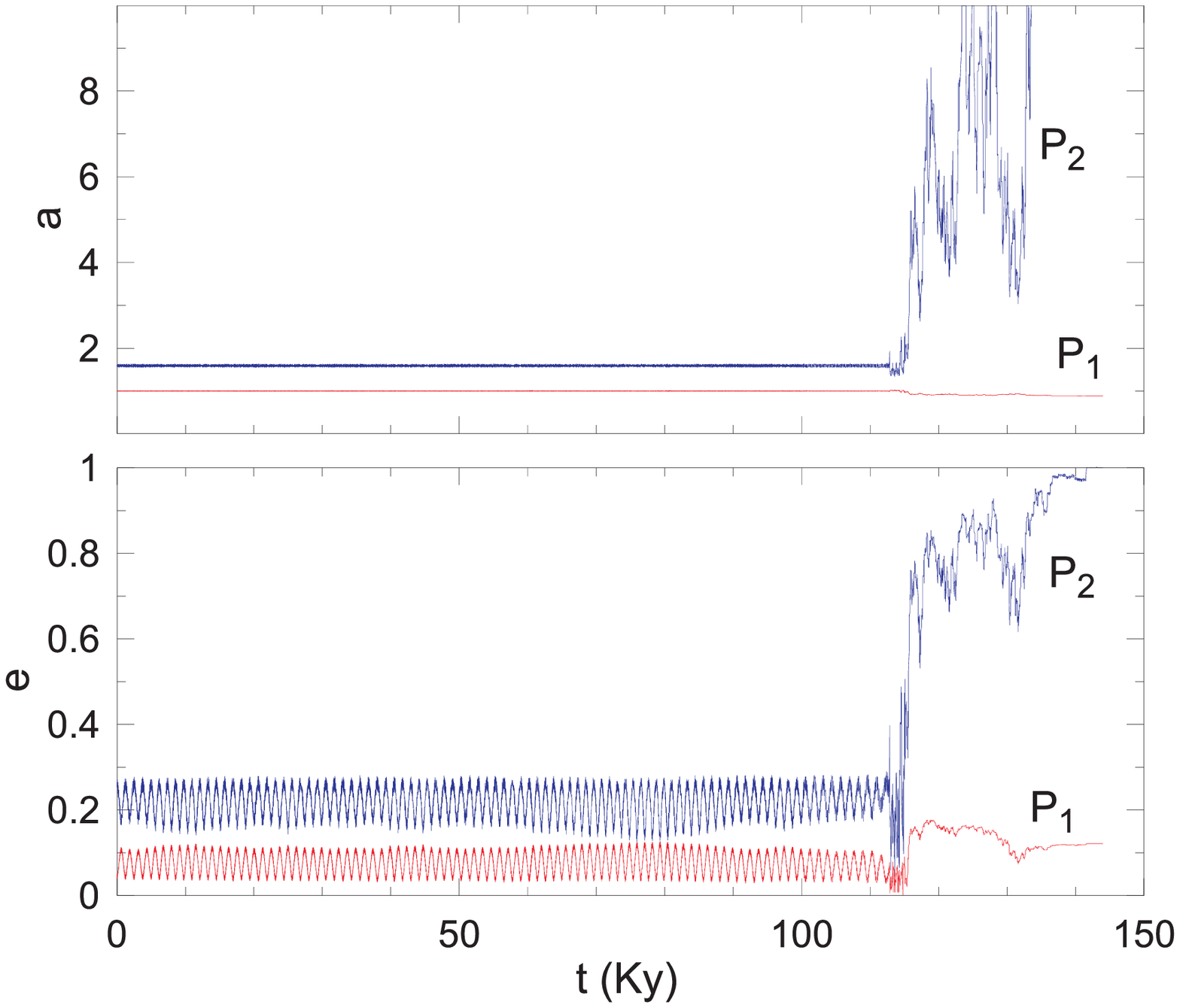}\\ 
\textnormal{(a)} & \textnormal{(b)} \\
\end{array}
$$ 
\caption{The evolution of semimajor axis and eccentricity of the planets $P_1$ and $P_2$ along a chaotic orbit with initial conditions same to those in the DS map of figure \ref{FigRESmaps}a and initial eccentricities (a) $e_1=0.05$, $e_2=0.3$  (b) $e_1=0.05$, $e_2=0.26$. }  
\label{FigEvolae} 
\end{figure} 

Since the system is of more than two degrees of freedom, invariant tori are not boundaries for the chaotic regimes in phase space and fast or slow diffusion of chaotic motion is possible (see e.g. Lichtenberg and Lieberman, 1983). Also, if we take into account that in the GTBP zero velocity curves do not exist to bound the planetary orbits (Marchal and Bozis, 1982), then we may conjecture that chaotic motion leads to the escape of a planet after a long term evolution.  This has been established after extensive numerical simulations of the GTBP (see Valtonen and Karttunen 2006, and references therein). In the following we discuss the particular case of a planar two-planet system.
  
In the dynamical maps presented in the previous subsections, we observed wide chaotic regions with strongly chaotic motion. Many numerical integrations of orbits with initial conditions inside these regions showed that, during the evolution, close encounters between planets occur and the less massive planet is scattered to orbits with larger eccentricities. The continued irregular evolution results in a sequence of close encounters that ejects the planet to large distances and finally to escape. We classify the evolution as escape when a planet that starts from a distance of order 1 AU, moves to a distance larger that 1000 AU with eccentricity $e>1$. In these numerical integrations we set the accuracy in the BS integrator to 14 digits and we use the regularized equations during close encounters. 

In Fig. \ref{FigEvolRmin} we present the evolution of the planetary distance for an orbit starting from the strongly chaotic (light colored) region appearing in the map of dynamical stability of Fig. \ref{FigRESmaps}a at $(e_1,e_2)=(0.05,0.3)$. The first close encounter appears after about 8 Ky. A sequence of close encounters follows and after 20 Ky scattering of the planet $P_2$ to large distances is observed. Finally the planet escapes at about 30 Ky. In Fig. \ref{FigEvolae}a we present the evolution of the semimajor axis and the eccentricity. After the first close encounter the eccentricity and the semimajor axis of $P_2$ show a jump to higher values and their variation in time becomes very irregular. The orbit of the heavier planet $P_1$  shows also the same irregularity, but the variation of its orbital elements is quite smaller. Finally we get $e_2>1$ and for the remaining planet $e_1=0.13$, $a_1=0.89$.

A second example of chaotic evolution is presented in Fig. \ref{FigEvolae}b. Now we start from the point $(e_1,e_2)=(0.05,0.26)$ which is closer to the regular (dark colored) region of the DS map of Fig. \ref{FigRESmaps}a. Up to 120 Ky the evolution of the semimajor axis and eccentricity of both planets seems vary regular. Afterwards chaotic evolution appears and a sequence of close encounters takes place, similar to the previous case. Finally planet $P_2$ escapes and the remaining planet $P_1$ moves on an elliptic orbit with $e_1=0.12$ and $a_1=0.89$. By considering initial conditions that correspond to chaotic motion but are closer to the regime of regular motion, we obtain similar planetary destabilization, which, however, occurs after a longer time span. For example, by considering the same initial conditions as above but for $e_2=0.25$ and $e_2=0.23$  we found that escape appears at 0.9My and 10My, respectively. However we should remark that the escape time is very sensitive to the numerical integration accuracy.   

Chaos is a necessary but not sufficient condition for planetary ejection over reasonable time spans (e.g. MS lifetimes). If we consider the initial conditions in a narrow chaotic zone in the maps of dynamical stability and far from the wide chaotic sea, then chaos may remain bounded throughout long-term evolution. In other words, consider a system with initial conditions in the chaotic region comprising small eccentricities ($e_i<0.15$) of the DS map of Fig.\ref{FigRESmaps}a.  Although this system evolves apparently irregularly, up until 100My the system is stable in the sense that no ejection or collision occurs.  The existence of bounded chaos for orbits which are Hill stable is also noted by \cite{Gladman93}. 

\section{Evolution under mass loss from the star and gravitational interaction between the planets}\label{mass-grav}

Here we present our main results.  We analyze how mass loss could couple with the planets' mutual interaction to cause one of them to escape.  In the following study we start with a two-planet system, which is trapped in a stable (resonant or non-resonant) configuration. Then we assume that the star begins to lose mass in an isotropic way, but in a manner where the percentage of the mass lost and the rate of mass loss are not large enough to cause the direct ejection of the planet. However, as we showed in Section 2, even in these cases the orbital elements of the planetary orbits are affected. So the system migrates in phase space and after the end of the mass loss process it can be found inside a chaotic region. From that point on, the chaotic evolution of the system suffers from close encounters, allowing for the possible ejection of a planet (see section \ref{chaotic}). 

We consider a linear rate of mass loss, $m=m_0-\alpha t$, which takes place up to time $t_\ell=\beta m_0/\alpha$ where the mass of star is reduced to  $m=(1-\beta)m_0$ (see section 2).  In all the figures showing the evolution of the system, the inner planet has the subscript ``1'' and is represented by a blue line and the outer planet has the subscript ``2'' and is represented by a red line.

\begin{figure}[tb]
$$
\begin{array}{cc}
\includegraphics[width=6cm]{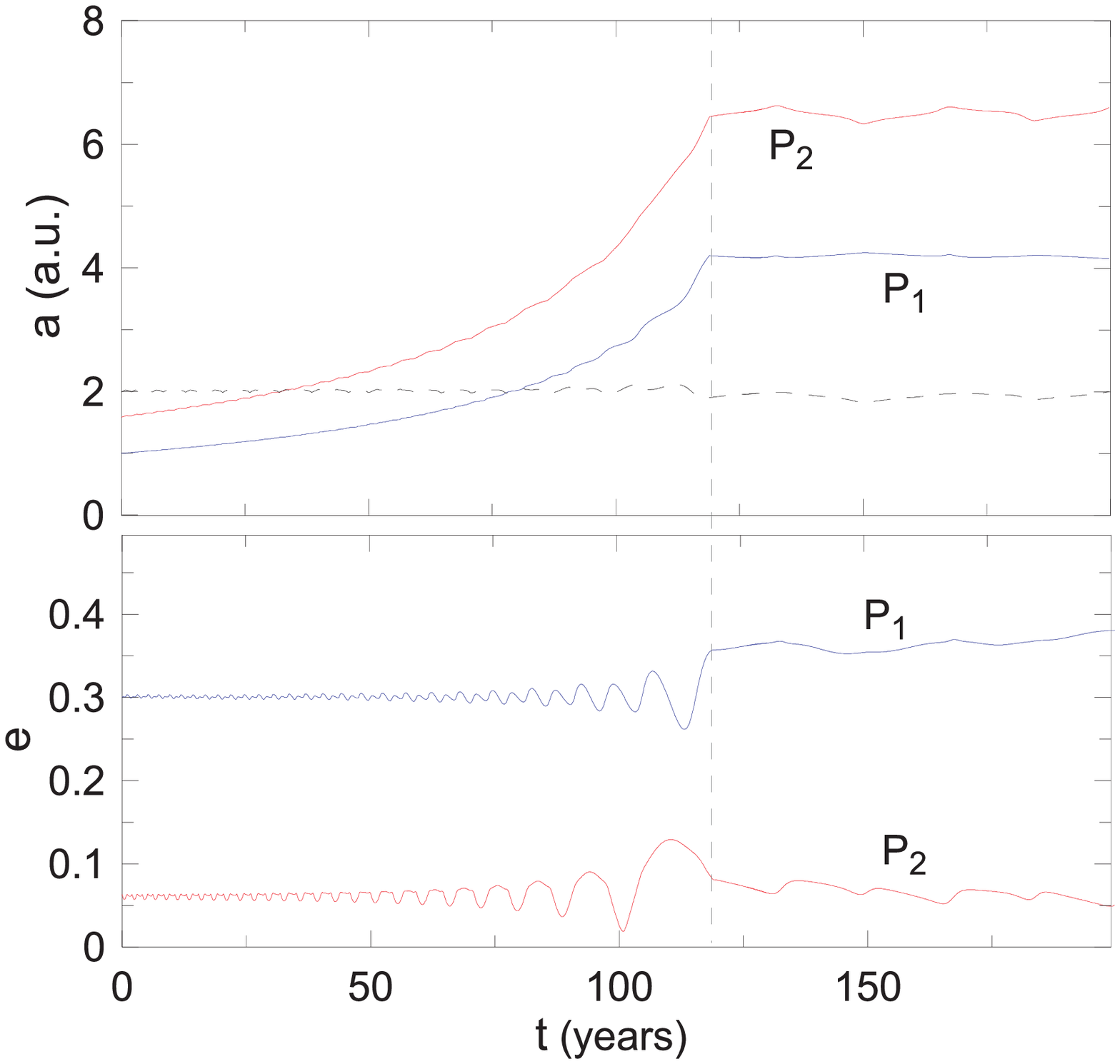} & \includegraphics[width=6cm]{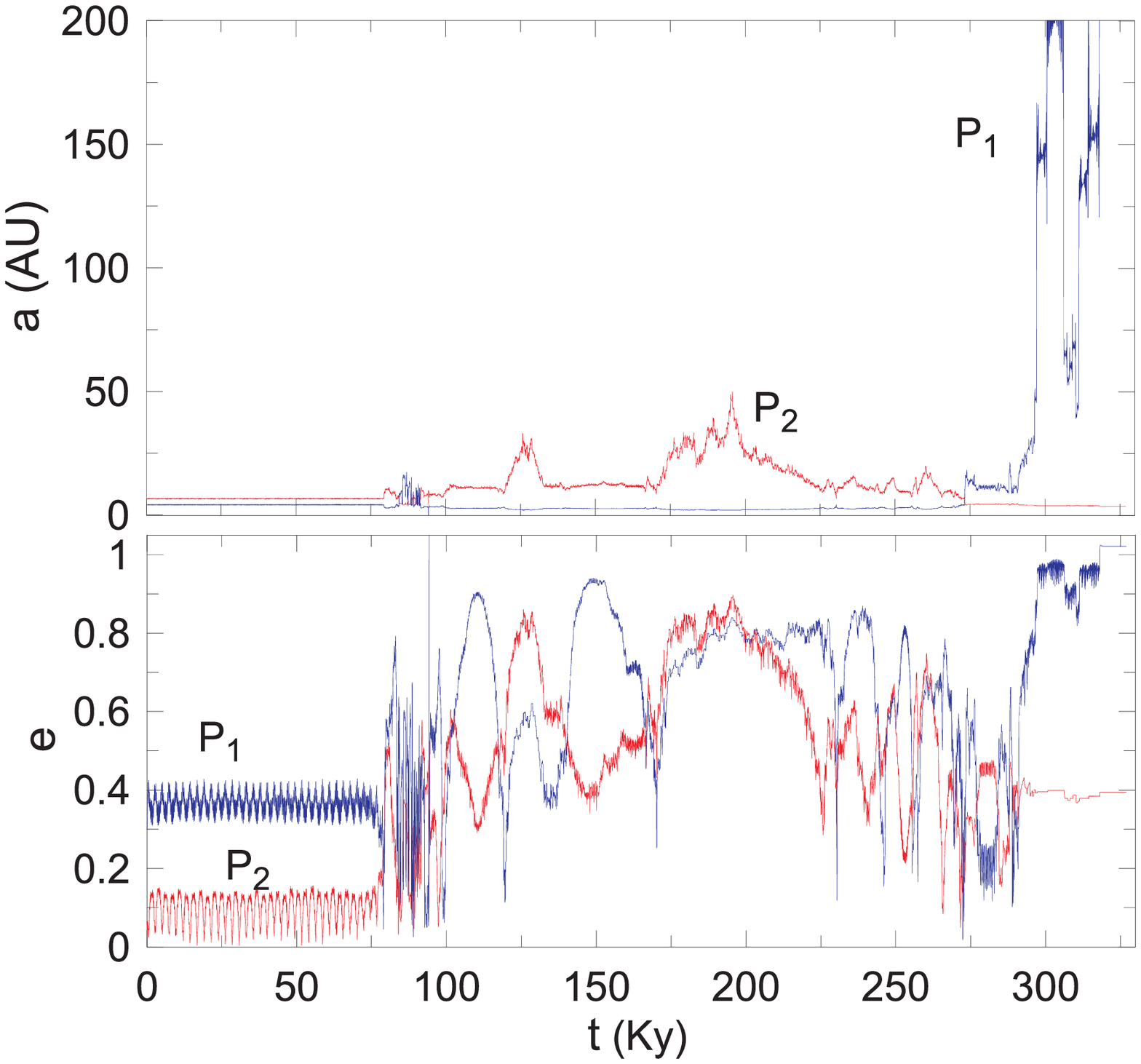} \\ 
\textnormal{(a)} & \textnormal{(b)}
\end{array}
$$ 
\caption{The evolution of the planetary semimajor axes and the eccentricities for the periodic orbit of Fig. \ref{FigPOmaps}a when it is affected by the star mass loss with $\beta=75\%$ and $\alpha=10^{-3}$. In panel (a) the first stage of the evolution is presented where the vertical dashed line indicates the time $t_\ell$ where the mass loss ceases. In panel (b) the continued evolution is presented until the escape of the planet $P_1$.}  
 \label{FigVMIpo} 
\end{figure}

\subsection{A fiducial example}

Here, the initial total mass of the system is set to one solar mass, namely $m_0\approx 1$. As an example we consider 
the stable 2/1 resonant periodic orbit given in Fig. \ref{FigPOmaps}a and we apply a star mass loss $\beta=75\%$ and rate $\alpha=10^{-3}$. The evolution of the planetary semimajor axes and eccentricities for the first years of the mass loss is shown in Fig. \ref{FigVMIpo}a. 
The semimajor axes increase for both planets approximately according to the adiabatic estimate given by Eq. (\ref{SemiAxisEvol}), but the ratio of mean motions $n_1/n_2$, represented by the horizontal dashed line, is almost constant, as it is suggested by Eq. (\ref{ResEvol}). After the end of the mass loss process, which ceases at 120 years, the eccentricities have increased slightly but the mass of the star has decreased significantly. Such new conditions corresponds now to a chaotic orbit as it becomes evident in Fig. \ref{FigVMIpo}b. We observe that the eccentricities incur well-bounded but weakly chaotic oscillations up to 70 Ky. Then the system enters a strongly chaotic region and the eccentricities show large and irregular variations up to about 290 Ky. In this interval, the variation of the semimajor axis of the planet $P_2$, which is the heavier one, is significant. After this time interval the semimajor axis and the eccentricity of the planet $P_1$ increase rapidly and finally the planet is ejected. It is obvious that if we ignore the gravitational interactions between the planets, the orbits of both planets would be constant ellipses with orbital elements attained at $t=t_\ell$.  

\begin{figure}
\centering
\includegraphics[width=8cm]{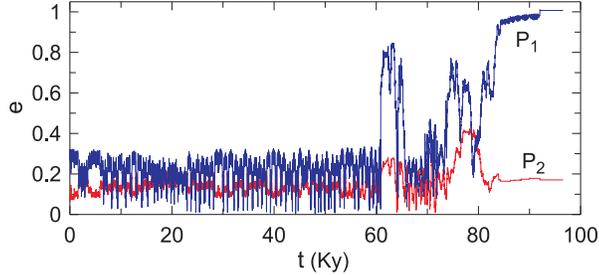} 
\caption{The eccentricity evolution of an orbit located near the periodic orbit of the DS map of Fig. \ref{FigPOmaps}a. The initial conditions are those of the periodic orbit except that we set $\omega_2=40^\circ$ and the mass loss parameters are $\alpha=10^{-4}$ and $\beta=50\%$.}
\label{FigVMIpoF}
\end{figure}

The above is a typical example of destabilisation after stellar mass loss. The same results are obtained for the periodic orbits in panels (b) and (c) of Fig. \ref{FigPOmaps}. As we mentioned in section 3, stable periodic orbits are in the centers of islands of stability. If we consider initial conditions far from the periodic orbit and near a chaotic region, destabilization can happen for smaller values of the parameters $\alpha$ and $\beta$. An example is shown in Fig. \ref{FigVMIpoF} where we use the same initial conditions except that $\omega_2$= $40^\circ$, which means that the orbit is located at the point $(\theta_1,\Delta\omega)$=$(-80^\circ,40^\circ)$ of the DS map of Fig. \ref{FigPOmaps}a. Now the system is destabilized for $\alpha=10^{-4}$ and $\beta=50\%$. 

\subsection{Maps of destabilization}
\begin{figure*}
\includegraphics[width=15cm]{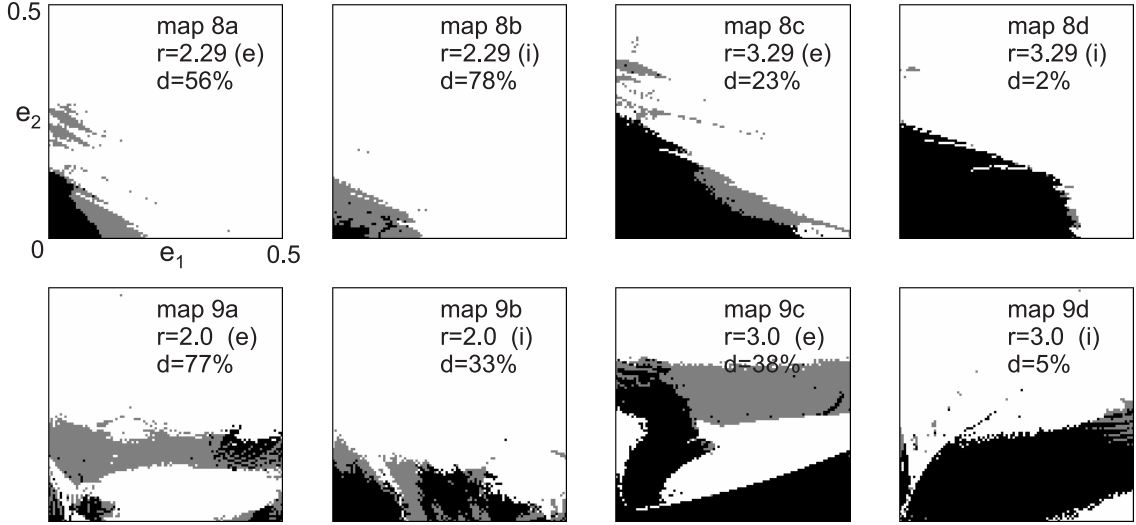} 
\caption{Maps which show the orbits that remain stable or destabilized after the mass loss of the star ($\alpha=10^{-4}$, $\beta=75\%$). The maps correspond to the DS maps of figures \ref{FigNRmaps} and \ref{FigRESmaps}, as it is indicated in each one of the maps. The ratio of mean motion $r$ is also indicated while the index (i) or (e) indicates that the less masive planet is the inner or the external one, respectively. The black areas indicate the orbits that continue to evolve regularly after the effect of the star mass loss. The gray area indicates the stable orbits that became chaotic. The percentage of the orbits that became chaotic is indicated by the index $d$.}  
 \label{FigVMmaps} 
\end{figure*}

In section \ref{secDSM} we presented DS maps on the plane of eccentricities, which show the distribution of regular and chaotic orbits when the mass of the star is constant. According to the mechanism described above, the regular orbits can become chaotic if stellar mass loss takes place. In order to make a more extensive study of the possibility of planetary destabilization, we consider the initial conditions of the regular orbits in the DS maps  of figures \ref{FigNRmaps} and \ref{FigRESmaps} and we follow the evolution of the system by considering star mass loss with $\alpha=10^{-4}$ and $\beta=75\%$. For each orbit we compute the LCN and thus we determine which of the orbits become chaotic after the stellar mass loss. We assume that if LCN is greater than $5 \times 10^{-5}$, then the evolution is chaotic. The results are presented in Fig. \ref{FigVMmaps}. The black shaded areas denote the orbits which remain regular after the mass loss process. The gray shaded area denotes the regions where the initially regular orbits become chaotic.

For each map we indicate the percentage $d$ of the orbits that become chaotic. In the resonant case with $r=2$ and the nonresonant one with $r=2.29$, we find that a large fraction of orbits is destabilised. In contrast, at the $3/1$ resonance and for the nonresonant case $r=3.29$, the majority of the orbits remain regular, especially when the outer planet is the less massive one. We note that in the first cases ($r=2$ and $r=2.29$) the planets are closer to each other and their gravitational interaction is stronger. Thus, we may claim that planetary orbits are more affected by stellar mass loss as the planetary interactions becomes stronger.  This important claim is not directly assessed in \cite{debsig2002} nor \cite{veretal12}, as the simulations in those studies are not set up to address this issue.  Further, in those studies, the planets are of the same mass and the outer planet is the one that escapes. In our integrations the planet which escapes is always the lighter one (see also \cite{dunclis98}).

\subsection{Realistic cases}

\begin{table}
 \centering
 \begin{minipage}{180mm}
  \caption{Thermally Pulsing Asymptotic Giant Branch Properties}
  \begin{tabular}{@{}cccc@{}}
  \hline
   Progenitor Mass & $\beta$ & $t_\ell$/(ky) & $\alpha$ \\
 \hline
 $8M_{\odot}$ & 77.8\% & 78.0 & $1.27 \times 10^{-5}$ \\[1pt]
 $7M_{\odot}$ & 78.5\% & 81.1 & $1.08 \times 10^{-5}$ \\[1pt]
 $6M_{\odot}$ & 78.8\% & 89.1 & $8.44 \times 10^{-6}$ \\[1pt]
 $5M_{\odot}$ & 78.0\% & 98.7 & $6.26 \times 10^{-6}$ \\[1pt]
 $4M_{\odot}$ & 76.3\% & 108.2 & $4.50 \times 10^{-6}$ \\[1pt]
 $3M_{\odot}$ & 73.3\% & 184.6 & $1.90 \times 10^{-6}$ \\[1pt]
 $2M_{\odot}$ & 65.5\% & 272.1 & $7.64 \times 10^{-7}$ \\[1pt]
\hline
\end{tabular}
\end{minipage}
\end{table}

Although this study is focused on the dynamical properties of the general three-body problem with mass loss, we
can relate the results to real systems.  Doing so helps us determine in what cases will significant
mass loss affect planet-planet scattering in a non-adiabatic manner.

In order to make this relation, we use the {\tt SSE} ({\it Single Star Evolution}) code \cite{huretal2000}
to create stellar evolutionary tracks.  For Solar metalicity stars, and a Reimers mass loss coefficient
of $0.5$, we trace the time evolution of stars with the following progenitor masses: $1M_{\odot}$,
$2M_{\odot}$,$3M_{\odot}$,$4M_{\odot}$,$5M_{\odot}$,$6M_{\odot}$,$7M_{\odot}$ and $8M_{\odot}$.  We find that
for all cases except the $1M_{\odot}$ case, the vast majority of the mass is lost on the Thermally Pulsing
Asymptotic Giant Branch (TPAGB) phase.  This phase, which is typically one of the shortest in duration,
also typically features the greatest mass loss rates.  This combination is well-suited for observing
the type of phenomenon seen in the fiducial simulations.  Therefore, we neglect the other phases
of evolution and assume that all the most lost occurs during the TPAGB.

We summarize the output of {\tt SSE}, which includes $\beta$ and $t_\ell$ (for the TPAGB phase only) 
in Table 1.  Then we remove the assumption that $m_0\approx 1$ and compute $\alpha$ and record that 
value as well in the last column.  The table
suggests that adopting $\alpha \sim 10^{-5}$ may reflect real systems with progenitor stellar masses 
of $6M_{\odot} - 8M_{\odot}$.

\begin{figure}[tb] 
$$
\begin{array}{cc}
\includegraphics[width=7cm]{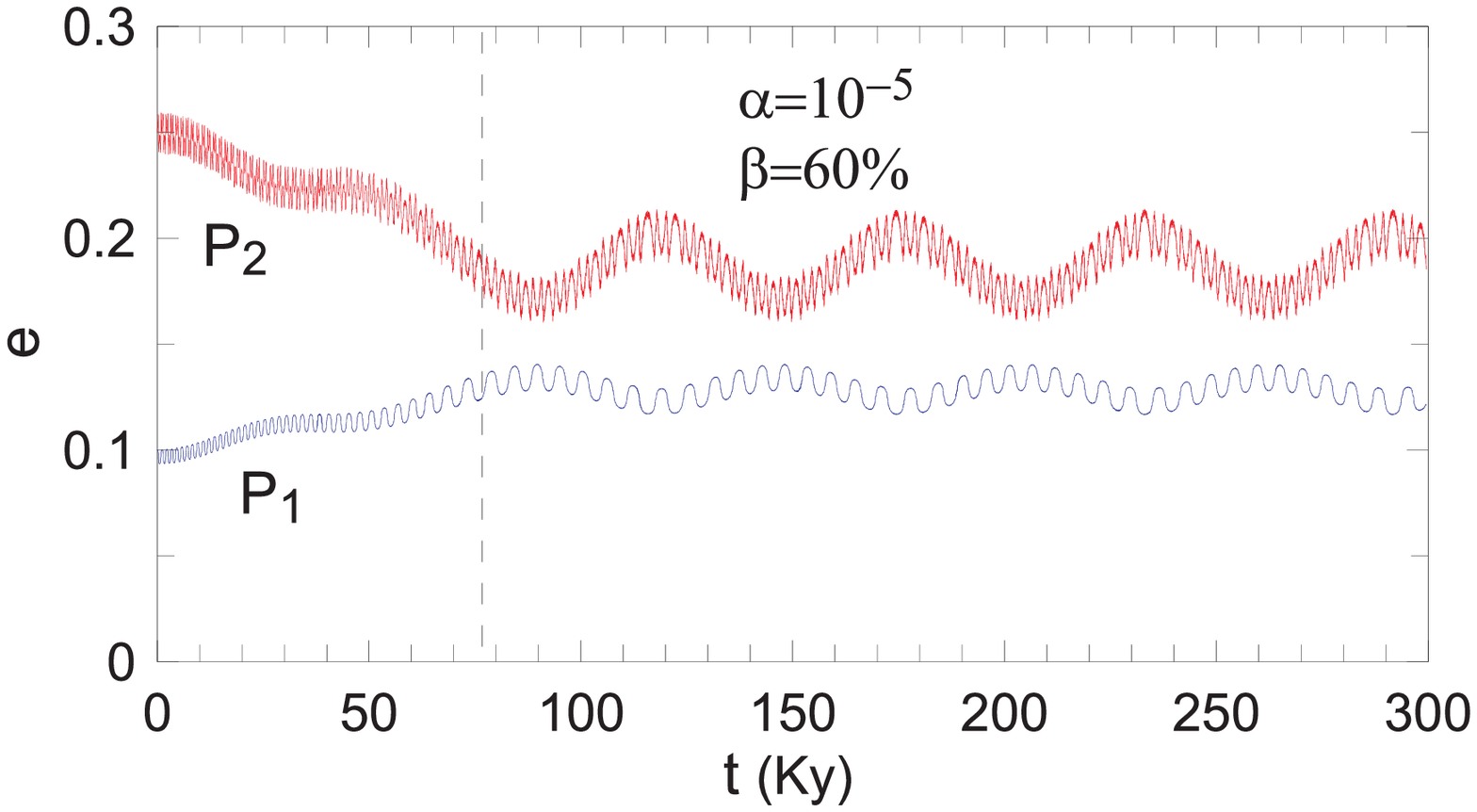} & \includegraphics[width=7cm]{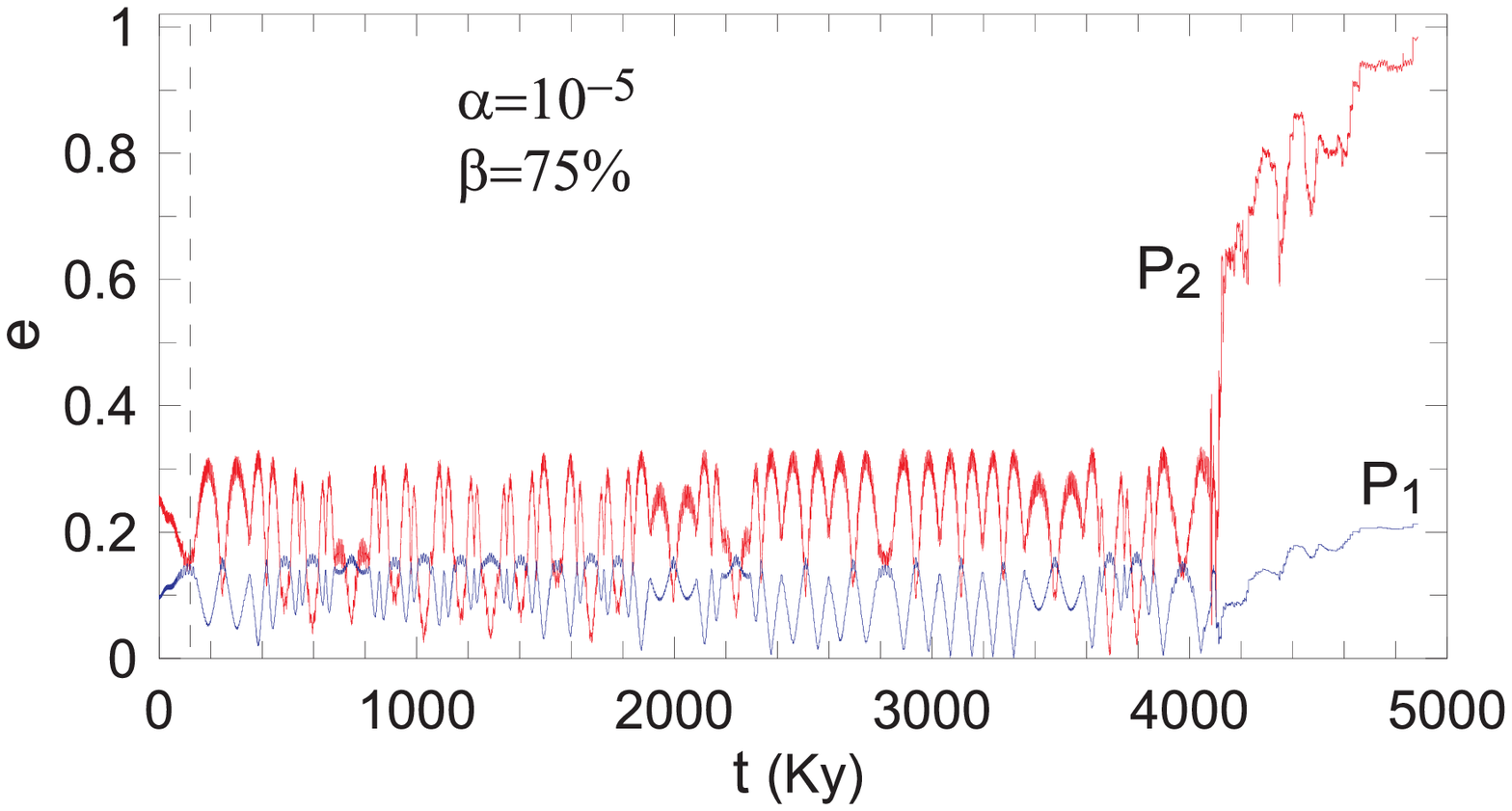}\\ 
\textnormal{(a)} & \textnormal{(b)} \\
\end{array}
$$ 
\caption{The eccentricity evolution for a 2/1 resonant planetary system with $m_0=8M_\odot$ and $m_1=3M_J$ and $m_2=0.6M_J$. The star loses mass linearly with rate $\alpha=10^{-5}$ and $\beta=60\%$ and $\beta=75\%$ (panel (a) and (b), respectively). The dashed vertical line indicates the $t_\ell$ value, where the mass loss ends. The initial conditions are $a_1=10.0$ and $a_2=15.97$, $e_1=0.1$, $e_2=0.25$, $\omega_1=0^\circ$, $\omega_2=180^\circ$, $M_1=M_2=0^0$. }  
\label{FigEvolm8} 
\end{figure}

In Fig. \ref{FigEvolm8}, we show a typical example of the evolution of such a multi-planet system with a star of mass $8M_{\odot}$ and two planets with $m_1=3M_J$ and $m_2=0.6M_J$. The planets are initially located in the 2/1 resonance with initial eccentricities $e_1=0.1$ and $e_2=0.25$. The numerical integration shows long term stability if the star does not lose mass. In particular, the eccentricities oscillate regularly about their initial values.  If we introduce a mass loss process with a rate $\alpha=10^{-5}$ and $\beta=60\%$ then we obtain the evolution given in Fig. \ref{FigEvolm8}a. In the time interval, $t<t_\ell$ the eccentricities show an adiabatic variation besides their fast oscillations. When the mass loss ceases, the system continues to evolve regularly with the eccentricities oscillating (with fast and slow components) about the values 0.13 and 0.19 for the inner and the outer planet, respectively.

The evolution becomes very different if we consider a larger amount of mass loss. Particularly for $\beta=75\%$ we get the evolution shown in Fig. \ref{FigEvolm8}b. Now, at $t=t_\ell$ the planetary system has entered a chaotic region and its evolution for $t>t_\ell$ is evidently chaotic. After 4.1~My close planetary encounters destabilize the system and the mutual gravitational attraction of the planets forces the outer, lighter, planet to be ejected from the system after about 5~My. 

\begin{figure}[tb]
\centering
\includegraphics[width=8cm]{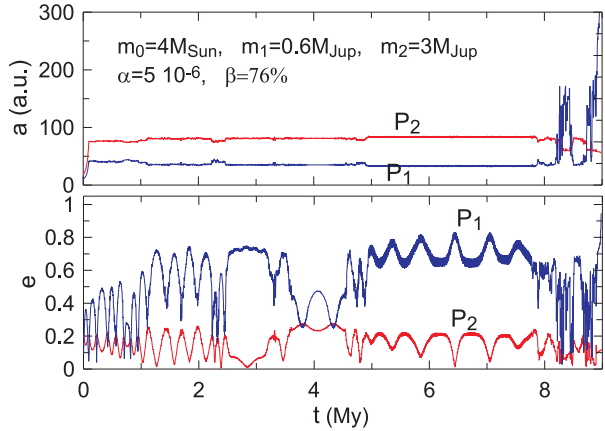} 
\caption{The semimajor axis and eccentricity evolution of a system with $m_0=4M_\odot$, $m_1=0.6M_J$ and $m_2=3M_J$. The star loses mass linearly with rate $\alpha=5 \times 10^{-6}$ and $\beta=76\%$. The initial conditions are $a_1=10.0$ and $a_2=18.0$, $e_1=0.1$, $e_2=0.2$, $\omega_1=0^\circ$, $\omega_2=180^\circ$, $M_1=M_2=0^0$. }
\label{FigEvolm4}
\end{figure}

 Evolution in resonance is not a necessary condition for destabilization. Figure \ref{FigEvolm4} shows the evolution of a non-resonant system ($n_1/n_2\simeq 2.415$) which consist of a 4$M_\odot$ star and two planets with $m_1=0.6M_J$, $m_2=3M_J$, $a_1=10.0$, $a_2=18.0$, $e_1=0.1$ and $e_2=0.2$.  Note that in this case the inner planet is the smaller one. Now we consider realistic values of $\beta=76\%$ and $\alpha=5 \times 10^{-6}$ (see Table 1). Under the stellar mass loss the semimajor axes increase to $a_1=40~AU$ and $a_2=75~AU$. 

The system shows some strong irregularities in the evolution of eccentricities and the orbit of the inner planet becomes very eccentric. Particularly, in the time interval $5<t<8~My$ the eccentricity of the outer planet oscillates about its initial value but the orbit of the inner planet shows eccentricity values in the interval $0.6<e_1<0.8$. For $t>8~My$ close planetary approaches force the inner and lighter planet to become (temporarily) the outer one in the system. The evolution becomes very irregular and, finally, the lighter planet is ejected out to a distance larger than 1000~AU. We have also performed integrations (up to $100$ My) indicating that, without mass loss, this system is stable and regular.  Therefore, mass loss is the trigger for the instability in this 4$M_\odot$ case.
 
In general the numerical simulations show that the position of the system in phase space determines essentially its robustness to perturbations caused by stellar mass loss. Also, the planetary destabilization seems to be sensitive to the parameters $\alpha$ and $\beta$. Concerning the case of systems of more than two planets and also taking into account the results of \cite{dunclis98} and \cite{debsig2002}, we may claim that these systems may be more sensitive to small perturbations than 2-planet systems. As soon as more planets are included in a system, the regions of stability in phase space is reduced and instabilities are more likely to occur after a potential decrease of the star's mass.

\subsection{Link with Observed Exosystems}

Our study might help explain discrepancies in the observed populations of 2-planet systems on the MS with those beyond the MS.  To date, no planets orbiting white dwarfs have been confirmed, although white dwarfs are present as distant companions in known planetary systems (Gl 86: \cite{queetal2000,mugneu2005}; HD 27442: \cite{butetal2001,chaetal2006,ragetal2006,mugetal2007}; HD 147513: \cite{mayetal2004,mugneu2005,pordas1997}).  Alternatively, over 30 planets are known to orbit giant stars (see Table 6 of \cite{getetal2012}) and well over 600 are confirmed to orbit MS stars\footnote{See the Extrasolar Planet Encyclopedia at http://exoplanet.eu/}$^{,}$\footnote{See the Exoplanet Data Explorer at http://exoplanets.org/}.

Among the mechanisms invoked to explain the general lack of known post-MS planets are different modes of formation around giant stars \cite{currie2009} and direct engulfment into the star itself due to star-planet tides (e.g. \cite{villiv2007,villiv2009,noretal2010,kunetal2011,musvil2012,norspi2012}).  We are proposing a third mechanism: scattering in two-planet systems that is induced by mass loss.  The planets in these systems are far away enough from their parent star to remain unaffected by tidal effects.  A potential fourth mechanism, albeit one which will be complex to model, is two-planet scattering where the planets are close enough to the star to be affected by both mass loss and tides.  This represents a potentially relevant extension to this work given that at least one planet in the vast majority of currently observed 2-planet systems will likely be affected by tides during post-MS evolution.  Therefore understanding the fate of multi-planet systems crucially depends on effects such as the one we have modelled here.

\section{Conclusion}

We have explored properties of the three-body problem with mass loss.  An
astrophysical application for this work is two-planet post-MS systems. We proposed a mechanism, which combines the effects of stellar mass loss and mutual planetary gravitational interactions, that can destabilize a planetary system.  Consequently, the lighter planet escapes and becomes an orphan planet.
   
In particular, the coupling between mass loss and mutual interactions
between the planets is not well-understood and was previously largely unexplored.
We have identified the regions of phase space where the coupling is strong,
causing initially stable systems to remain stable during the mass loss but 
later exhibiting escape. The stellar mass loss may transfer the system from its initially stable region to a wide chaotic sea in the phase space, which is associated with the final stellar mass. In such a chaotic region escape of the lighter planet always occurs. Although we find this evolutionary 
sequence to be robust to small changes in initial conditions, we find that
the escape time is sensitive to these changes. However, it seems that chaotic evolution, which is detected by the maximal LCN during an integration interval up to $\sim 100$ ky, results in most cases to destabilization of the system in less than 100 My. The general phase space maps presented can be used to characterize a wide variety of real systems.

In the numerical simulations of our paper we used a constant rate of mass loss. However, we have obtained qualitatively similar results for an exponential rate of mass loss: Fig. 4 demonstrates excellent agreement between the constant mass loss model and the exponential model.  Also, we note that mass loss along the Asymptotic Giant Branch closely resembles either constant or exponential mass loss. 

We find that the upper-mass end of potential planetary systems that
will experience a white dwarf phase ($M \sim 4-8 M_{\odot}$)
is likely to exhibit the chaotic evolutionary behaviour seen here.  Just as importantly, 
lower-mass planetary systems, such as the currently-observed bound multi-planet exoplanet 
population, will experience adiabatic planet-planet interactions during post-MS
mass loss.  Claiming that the adiabatic approximation holds in these cases for the
three-body problem is useful, and will simplify the analytics for future 
post-MS three-body studies.

\vspace{0.5cm}
{\bf Acknowledgements}
We thank the referee for helpful suggestions.  This research was supported by the Grant of the A.U.TH Research Committee (Greece): ``Action C:Support of Basic Research'', contract No 87840.

\label{lastpage}

\end{document}